\documentclass[onecolumn,pra,amsmath,amssymb,showkeys]{revtex4}

\usepackage{amsmath}
\usepackage{graphicx}
\usepackage{graphics}
\usepackage{dcolumn}
\usepackage{bm}
\usepackage{hyperref}

\def\BEq{\begin{equation}}
\def\EEq{\end{equation}}
\def\BEqA{\begin{eqnarray}}
\def\EEqA{\end{eqnarray}}
\def\BEn{\begin{enumerate}}
\def\EEn{\end{enumerate}}
\def\BWT{\begin{widetext}}
\def\EWT{\end{widetext}}

\def\z{\zeta}

\begin{document}

\title{Feynman's method in chiral nanorod-based 
metamaterial nanoplasmonics}

\author{Andrei Galiautdinov}
\email{ag1@uga.edu}
 \affiliation{
Department of Physics and Astronomy, 
University of Georgia, Athens, GA 30602, USA}

\date{\today}
\begin{abstract}

We propose a theoretical approach to some of the nanorod-based metamaterial implementations that does not depend on macroscopic electrodynamics. 
The approach is motivated by the fact that in actual experiments the incident electromagnetic wave encounters a metamaterial structure which is planar in its shape, contains a layer or two of artificially created building blocks, and therefore cannot be regarded as a three-dimensional continuous medium. This leads to a theoretical framework in which the phenomenological concept of refractive index loses its principled meaning, and the deeper concept of scattering is taking center stage. Our proposal and its mathematical realization rely heavily on Feynman's explanation of the physical origin of the index of refraction and on his formula for the field of a plane of oscillating charges. We provide a complete proof of Feynman's formula, filling in some steps that were missing in the original derivation, and then generalize it to the case of a finite disk, which may be relevant to the actual experiments involving laser beams. We then show how the formula can be applied to metamaterial nanoplasmonics by considering some subtle interference effects in narrow laser beams striking metamaterial plates. The first two effects use a single layer of aligned plasmonic nanorods, while the third uses a single layer of gyrotropic elements that may conveniently be described by the celebrated Born-Kuhn oscillator model. The considered effects can potentially be used in the development of quality standards for various metamaterial devices.

\end{abstract}


\maketitle

\tableofcontents


\vskip10pt
{\it 
[This is the version of the article before peer review or editing, as submitted by an author to Physica Scripta. IOP Publishing Ltd is not responsible for any errors or omissions in this version of the manuscript or any version derived from it. The Version of Record is available online at: \href{https://doi.org/10.1088/1402-4896/ad0f64}{https://doi.org/10.1088/1402-4896/ad0f64}.]}

\newpage

\section{Introduction}

Maxwell's classical electrodynamics
\cite{maxwell1873treatise}, 
the pinnacle of 19th century theoretical physics, supplemented by Lorentz' 
electron theory of matter \cite{lorentz1892theorie, lorentz1906versuch} require 
for their mutual conceptual consistency the existence of three clearly distinguishable 
length scales \cite{landau1984electrodynamics,jackson1999classical}. 
Each scale comes equipped with its own characteristic distance 
(as well as the associated volume element) that has very 
specific physical meaning. The first, {\it microscopic} scale
operates at the level of the fundamental structural elements of a given medium 
(atoms, molecules, etc.), where, depending on the context, the characteristic 
distance, $d_1$, represents either the typical size of the fundamental element or 
the separation distance between two nearest such elements. The second, 
{\it intermediate} scale defines the size of the so-called ``physical infinitesimally 
small volume'' \cite{landau1984electrodynamics,mikhailov2019equations}
over which the macroscopic averaging of various field quantities has 
to be performed. Its characteristic distance, $d_2$, corresponds to the size of 
an ideal probe that could be used to measure the field in a given experimental 
situation \cite{lorentz1906versuch,zangwill2013modern}. 
The third, {\it macroscopic} scale, $d_3$, characterizes the macroscopic volume 
in which the {\it continuous} (that is, macroscopically averaged) fields are actually 
defined. The standard example is provided by a plane electromagnetic wave of 
wavelength $\lambda$ propagating in a transparent medium, in which case 
$d_1$ is the separation distance between the molecules, $d_3$ is on the order 
of the wavelength, and $d_2$ is some distance that satisfies the strong 
inequality, $d_1 \ll d_2 \ll d_3$.

Of the three scales mentioned above, the intermediate scale occupies a very 
peculiar position. Its characteristic distance has to be sufficiently small for 
the measurement process to provide the detailed description of the system 
at hand (say, provide the field values in the region which is several wavelengths 
long), and yet it has to be sufficiently large for the associated characteristic 
volume to contain a great number of microscopic structural elements for the 
averaging procedure to be meaningful. Oftentimes, the $d_2$ is not even 
explicitly mentioned or discussed, or its justification is sketchy and hand-waving, 
which makes the presence of the intermediate scale either ambiguous or 
not immediately clear. In spite of all these imperfections in its definition, 
whenever one deals with a medium that is regarded as continuous, the 
intermediate scale must always unreservedly be present. Its very existence, 
no matter how hypothetical, is absolutely indispensable for any ``legitimate'' 
use of continuous electrodynamics. This rather stringent requirement stems 
from the epistemological position that, since Maxwell's equations describe 
experimentally measurable quantities, there should exist a practically 
realizable method of their determination.

On the other hand, we know from everyday experience that too strict 
adherence to  operational definitions is not at all necessary for the productive 
development of a theory. Two very instructive examples immediately come 
to mind, although they are not directly related to the topic under consideration. 
The first of them is the discovery by Dirac 
\cite{dirac1981principles} of his famous wave equation for the electron in 
which there appeared some mysterious matrices that had no immediate 
physical interpretation (due to the presence of ``additional'' components 
and indirect relation to the velocity operator).
Another example is the experimental interpretation of the eigenvalues of 
the position operator for an electron in a hydrogen atom. To determine 
electron's position one has to strike the atom with highly energetic photons 
which would destroy the atom, a procedure similar to the one reported 
in \cite{stodolna2013hydrogen} in which the authors used photoionization 
and an electrostatic lens to directly observe the electron orbitals. 
So even though the system in question would be destroyed by the actual
measurement, one still uses the quantum mechanical language of wave functions 
and position operators to describe it.

When we turn our attention to metamaterial electrodynamics, we encounter 
a similar dichotomy. Often, metamaterial structures consist of just a few 
layers of artificially created building blocks and therefore cannot be regarded 
as three-dimensional continuous media for the purpose of macroscopic averaging. 
There is no concept of intermediate scale here, 
since the entire width of the system is comparable to the size of the 
``microscopic'' structural element. Nevertheless, when applied to 
such artificial noncontinuous structures, macroscopic electrodynamics 
(together with the standard procedures for retrieval of effective 
material parameters \cite{smith2005electromagnetic,kwon2008material,
wang2009chiral,cai2010optical,tassin2012effective})
seems to be working remarkably well in predicting system's electromagnetic 
properties
\cite{sihvola1991bi,lindell1994electromagnetic,vinogradov2002form,
maier2007plasmonics,wang2009chiral,schaferling2012tailoring,
yao2014plasmonic,goerlitzer2021beginner}; so much so that one 
can't help wondering if something deep is at play here underlying the 
theory's success. These considerations naturally lead to the question of why macroscopic 
electrodynamics relates so well to metamaterial systems in general, and 
nanoplasmonic ones in particular. It seems 
intuitively obvious that at the fundamental level the concept of scattering 
should play a major role in answering this important question. One 
needs to decide, however, which specific approach to scattering to take 
and how much rigor to exercise. Much of past work involved accurate 
and detailed studies of scattering on individual structural elements, 
taking into account some nontrivial near-field effects, followed by 
careful statistical averaging to describe the corresponding effective 
medium that would mimic system's electromagnetic response 
(see, e.\ g., Refs.\ \cite{lagarkov1996electromagnetic, 
ivanov2012plasmonic, vinogradov2001electrodynamics, 
vinogradov2002electrodynamics, sarychev2007electrodynamics, 
klimov2014nanoplasmonics}).
Here we propose an approach --- somewhat less detailed, but also rooted 
in scattering and capable of providing some deep physical insight --- 
to systems in which individual plasmonic nanorods (or their simple 
combinations) play the role of fundamental structural elements. We 
are primarily interested in either single sheets of aligned nanorods or 
metamaterial plates made of corner-stacked nanorods, which we 
propose to model --- and this is our main idea --- as {\it planes of oscillating 
charges}. This approach, whose details will be elucidated below, goes 
back to Feynman and is based on his formula for the field of a plane of 
oscillating charges, to which we now turn.

\section{Derivation of Feynman's formula for the field of a plane of 
oscillating charges}

\subsection{Preliminary considerations}

We assume the reader is familiar with Feynman's semi-intuitive 
derivation of his formula for the field of a plane of oscillating charges 
(see Eqs.\ (\ref{eq:F(30.18)}) and (\ref{eq:INFINITEplane}) below) 
as presented in \cite{feynman2006feynman}. The derivation misses 
some important steps which we aim to fill in here first, before discussing 
some applications in nanoplasmonics.

The importance of Feynman's formula stems from its use in explaining 
the physical origin of the refractive index of a dielectric medium whose 
underlying structural elements (e.\ g., atoms) may conveniently be 
described by the Drude-Lorentz oscillator model. 
To show how light ``slows down'' in a dielectric material, 
Feynman (\cite{feynman2006feynman}, Ch.\ 31) 
treats the field of the transmitted electromagnetic wave as 
consisting of two parts: the original (incoming) wave and the 
radiation field of all harmonically driven charges comprising the material.
He then subdivides the charges into a great number of planar 
sheets and asks for the contribution by one such sheet, at which 
point the formula for the field of an oscillating plane naturally enters 
the discussion. In Feynman's analysis the dielectric material is not 
regarded as a medium {\it per se}, but as a large collection of charged 
particles sprinkled around and scattering light {\it in a vacuum}, with 
{\it all} the fields propagating with the {\it same} limiting speed, $c$.
The apparent reduction in that speed then turns out to be nothing but 
an interference effect, a mathematical fluke, mathematical curiosity.  

\begin{figure} [!ht]
\includegraphics[angle=0,width=0.6\linewidth]{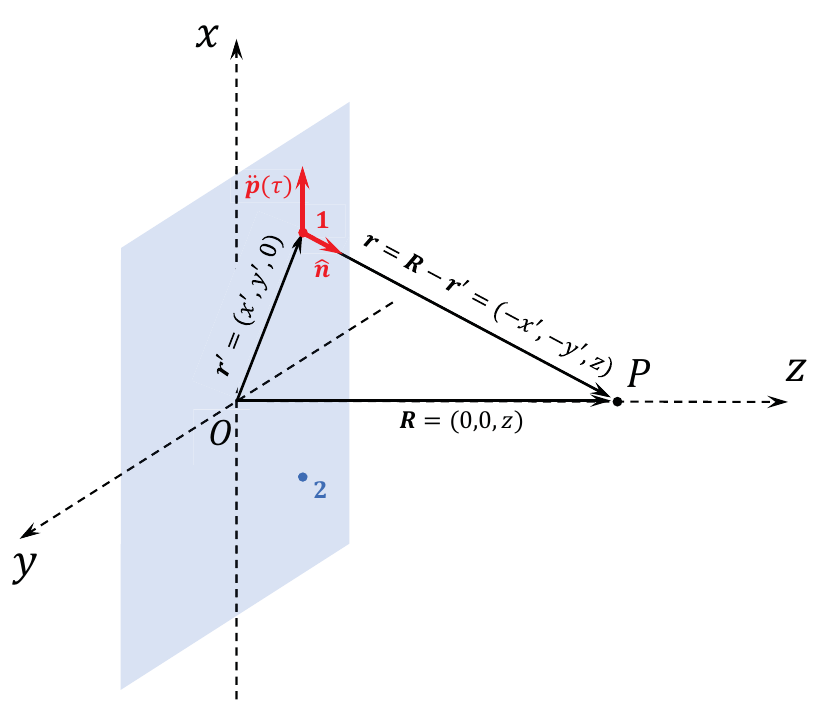}
\caption{ \label{fig:1}  
Geometry of the problem.
}
\end{figure}

Since each oscillating charge can be regarded as an oscillating dipole, 
${\bm p}$, we  begin our derivation of Feynman's formula with the 
{\it exact} expression for the field of a radiating point
dipole valid at any distance, $r$, regardless of the wavelength, $\lambda$ 
(Coulomb's constant dropped for notational simplicity and will be restored 
later; overdots indicate differentiation with respect to time),
\begin{align}
\label{eq:E(Pt)}
\bm{E}^{(\rm{dipole})}(P,t)
&=
\frac{3\left({\bm p}(\tau) \cdot \hat{\bm n}\right)\hat{\bm n} 
- {\bm p}(\tau) }
{r^3}
+
\frac{3\left(\dot{\bm p}(\tau) \cdot \hat{\bm n}\right)\hat{\bm n} 
- \dot{\bm p}(\tau) }
{c r^2}
+
\frac{\left(\ddot{\bm p}(\tau) \cdot \hat{\bm n}\right)\hat{\bm n} 
- \ddot{\bm p}(\tau) }
{c^2 r},
\quad
\tau = t-\frac{r}{c}\,.
\end{align}

According to (\ref{eq:E(Pt)}), in the far zone, the electric field vector 
$\bm{E}(P,t)$ evaluated at the observation point $P=(x,y,z)$ at time $t$  
is orthogonal to vector $\bm{r}$ connecting the dipole to $P$ 
and lies in the plane spanned by $\bm{r}$ and $\ddot{\bm p}(\tau)$, evaluated 
at the retarded time $\tau$ (with similar interpretations for the near and induction terms). 
In our problem all dipoles are assumed to be coherently oscillating in 
the $xy$-plane in the direction parallel to the $x$-axis, as shown in 
Fig.\ \ref{fig:1}. 
Correspondingly, for points $P=(0,0,z)$ located on the $z$-axis, 
all relevant vectors have components as indicated in that figure, with
${\bm p}(\tau)=({p}_x(\tau),0,0)$,
$\ddot{\bm p}(\tau)=(\ddot{p}_x(\tau),0,0)$, and
\BEq
\hat{\bm n} 
=\frac{\bm{r}}{r}
=\frac{(-x',-y',z)}{\sqrt{x'^2+y'^2+z^2}}\,.
\EEq

\begin{figure} [!ht]
\includegraphics[angle=0,width=0.5\linewidth]{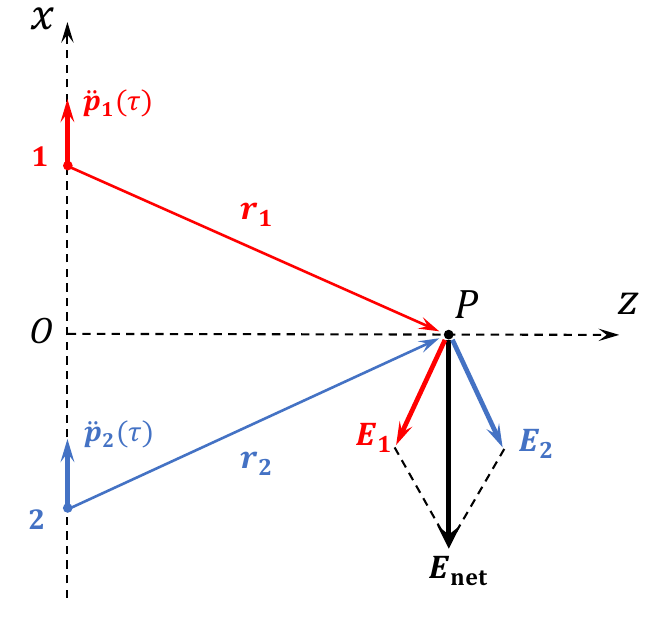}
\caption{ \label{fig:2}  
The far field of two symmetrically positioned dipoles has no $y$ 
and $z$ components.
Similar argument applies to the near and induction fields.
}
\end{figure}

Substitution into (\ref{eq:E(Pt)}) gives,
\begin{align}
\bm{E}^{(\rm{dipole})}(z,t)
&=
{p}_x(\tau)
\frac{\Bigg(2x'^2-y'^2-z^2, \, 3x'y', \, -3x'z\Bigg)}
{\left(x'^2+y'^2+z^2\right)^{5/2}}
\nonumber \\
& \quad
+
\frac{\dot{p}_x(\tau)}{c}
\frac{\Bigg(2x'^2-y'^2-z^2, \, 3x'y', \, -3x'z\Bigg)}
{\left(x'^2+y'^2+z^2\right)^{2}}
\nonumber \\
& \quad
+
\frac{\ddot{p}_x(\tau)}{c^2}
\frac{\Bigg(-y'^2-z^2, \, x'y', \, -x'z\Bigg)}
{\left(x'^2+y'^2+z^2\right)^{3/2}},
\end{align}
which shows that the $y$ and $z$ components of the electric field 
change sign under the $x' \leftrightarrow -x'$ transformation. This 
means that they vanish for any distribution of charges symmetric with 
respect to the $y$-axis, which can be understood intuitively by sketching 
the net field due to a pair of identical dipoles positioned symmetrically 
relative to the $y$-axis, with one directly above the other (Fig.\ \ref{fig:2}). 

Therefore, only the $x$ component of the field,
\begin{align}
\label{eq:Ex(Pt)explicit}
{E}^{(\rm{dipole})}_x(z,t)
&=
{p}_x(\tau)
\frac{2x'^2-y'^2-z^2}{\left(x'^2+y'^2+z^2\right)^{5/2}}
+
\frac{\dot{p}_x(\tau)}{c}
\frac{2x'^2-y'^2-z^2}{\left(x'^2+y'^2+z^2\right)^{2}}
-\frac{\ddot{p}_x(\tau)}{c^2} 
\frac{y'^2+z^2}{\left(x'^2+y'^2+z^2\right)^{3/2}},
\end{align}
needs to be kept in our derivation.

\begin{figure} [!ht]
\includegraphics[angle=0,width=0.5\linewidth]{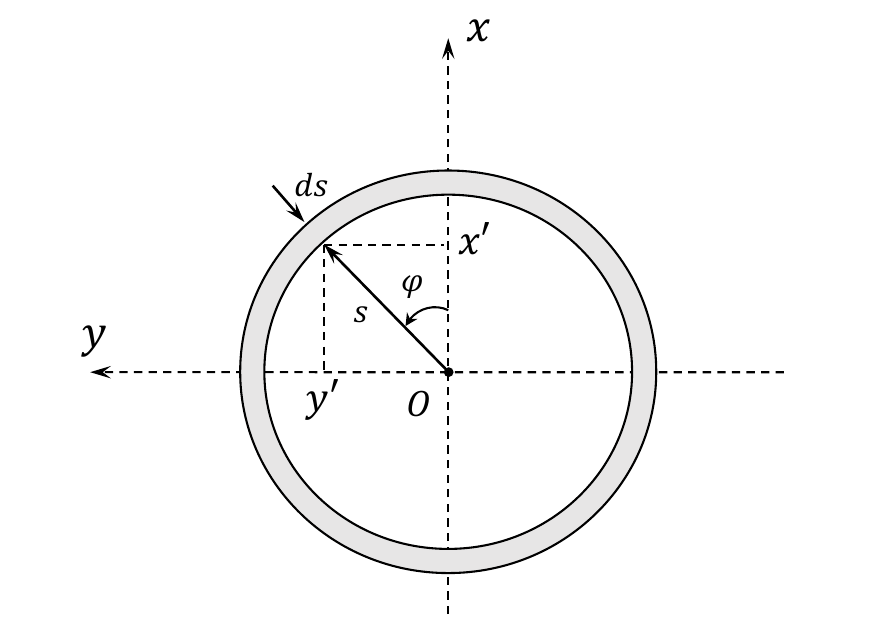}
\caption{ \label{fig:3}  
One of the infinitesimal rings used in the integration procedure.
}
\end{figure}

The calculation of the total 
field then proceeds by dividing the planar distribution into a 
series of infinitesimally thin concentric rings of variable radii $s$ 
and thickness $ds$, and then integrating (Fig.\ \ref{fig:3}).
For one such ring we have, on the basis of (\ref{eq:Ex(Pt)explicit}) 
and using $x'=s\cos \varphi$, $y'=s\sin \varphi$,
\begin{align}
\label{eq:Ex(Pt)explicit1}
{E}^{(\rm{ring})}_x(z,t)
&=
\eta \,{p}_x(\tau)
\int_{0}^{2\pi}
\frac{2s^2 \cos^2 \varphi -s^2 \sin^2 \varphi -z^2}
{\left(s^2+z^2\right)^{5/2}}
\, s \, d \varphi \, ds
\nonumber \\
&\quad
+
\frac{\eta \, \dot{p}_x(\tau)}{c}
\int_{0}^{2\pi}
\frac{2s^2 \cos^2 \varphi -s^2 \sin^2 \varphi -z^2}
{\left(s^2+z^2\right)^{2}}
\, s \, d \varphi \, ds
\nonumber \\
&\quad
-\frac{\eta \, \ddot{p}_x(\tau)}{c^2} 
\int_{0}^{2\pi}
\frac{s^2 \sin^2 \varphi +z^2}
{\left(s^2+z^2\right)^{3/2}} \, s \, d \varphi \, ds \, ,
\end{align}
where $\eta$ is the surface density (the number of dipoles per unit area).
Taking into account that
$\int_0^{2\pi}\cos^2 \varphi \, d\varphi 
= \int_0^{2\pi}\sin^2 \varphi \, d\varphi = \pi$,
$r^2=s^2+z^2$,
$sds = rdr$,
we get
\BWT
\begin{align}
\label{eq:Ex(Pt)explicit2}
{E}^{(\rm{ring})}_x(z,t)
&=
\pi \, \eta 
\left\{
{p}_x(\tau) \, \frac{1}{r^2}  \left(1-\frac{3z^2}{r^2}\right) 
+ \frac{\dot{p}_x(\tau)}{c} \, \frac{1}{r}  \left(1-\frac{3z^2}{r^2}\right) 
-\frac{\ddot{p}_x(\tau) }{c^2} \left(1+\frac{z^2}{r^2}\right) 
\right\}
\, dr \,,
\end{align}
\EWT
which, for harmonically oscillating charges with
\BEq
{p}_x(\tau) = q x_0 e^{-i\omega (t-r/c)},
\EEq
results in the {\it total} field,
\begin{align}
\label{eq:Ex(Pt)explicit3}
{E}_x(z,t)
&=
\pi \eta  q  x_0 \, e^{-i\omega t} \, I(z,a) \,,
\quad
I(z,a) &\equiv \int_z^a \, e^{i(\omega/c) r} 
\Bigg[
\frac{1}{r^2}  \left(1-\frac{3z^2}{r^2}\right) 
- \frac{i\omega}{c} \, \frac{1}{r}  \left(1-\frac{3z^2}{r^2}\right) 
+\frac{\omega^2}{c^2} \left(1+\frac{z^2}{r^2}\right) 
\Bigg]
\, dr \,.
\end{align}
In the above we used quantum mechanical sign convention for 
frequency $\omega$ and kept the upper limit of integration, $a$, 
finite with the intention of distinguishing  between two interesting 
cases: an infinite plane (in which case $a\rightarrow \infty$) and 
a disk (in which case $a$ remains finite). Before proceeding further, 
let us take a brief look at Feynman's own derivation.

\subsection{Brief review of Feynman's ``proof''}
\label{sec:feynman}

Feynman derived his result (see Sec.\ 30-7 in \cite{feynman2006feynman}),
\BEq
\label{eq:F(30.18)}
E_x(P) 
= 
2\pi \, \frac{\eta q}{c}\, i\omega x_0 e^{-i\omega (t-z/c)},
\EEq
by working in the far zone and using the approximate dipole radiation formula,
\BEq
E^{(\rm{dipole})}_x(P) \approx \frac{q}{c^2 r}\, \, \omega^2 x_0 \, 
e^{-i\omega (t-z/c)},
\EEq
which represents the field in dipole's equatorial plane only, thus ignoring  
the near zone, the induction zone, and the angular dependence due to 
mutual orientation of $\ddot{\bm p}(\tau)$ and $\hat{\bm{n}}$ in 
the far zone ({\it cf}.\ our Eq.\ (\ref{eq:Ex(Pt)explicit})). 
This leads to the expression for the total field,
\begin{align}
\label{eq:Ex(Pt)explicit3FEYNMAN}
E_x(P) 
=
2  \pi  \eta  q  x_0 \, e^{-i\omega t} \, 
\frac{\omega^2}{c^2} \int_z^a \, e^{i(\omega/c) r} \, dr ,
\end{align}
which differs substantially from our Eq.\ (\ref{eq:Ex(Pt)explicit3}). 
In the limit $a\rightarrow \infty$ the integral in 
(\ref{eq:Ex(Pt)explicit3FEYNMAN}) evaluates to
\BEq
\label{eq:F(30.16)}
\int_z^{\infty} \, e^{i(\omega/c) r} \, dr  
= 
-i\frac{c}{\omega}\left[e^{i(\omega/c)\infty} - e^{i(\omega/c)z} \right],
\EEq
which is not well defined.

\begin{figure} [!ht]
\includegraphics[angle=0,width=0.5\linewidth]{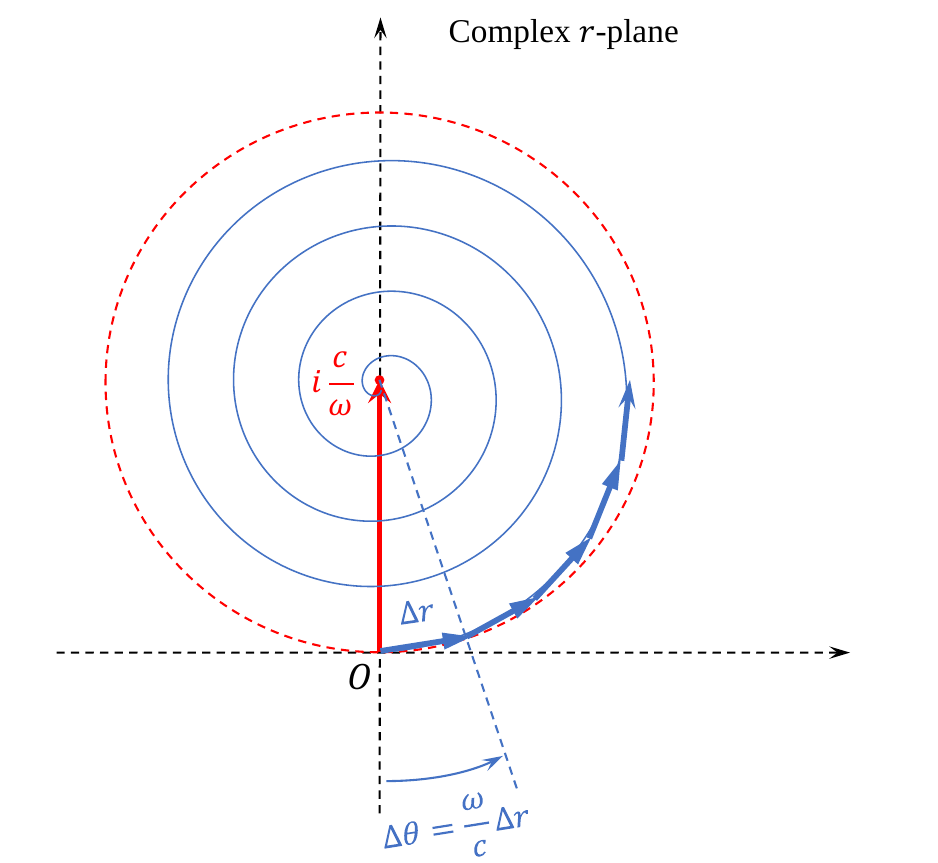}
\caption{ \label{fig:4}  
Feynman's approach to evaluating $\int_z^{\infty} \, \eta(r) \, 
e^{i(\omega/c) r} \, dr$ with the help of a radially tapered surface 
density $\eta(r)$ as given in Eq.\ (\ref{eq:integration_a_la_Feynman}).
}
\end{figure}

To make sense of the infinite exponent, Feynman regularizes 
(\ref{eq:Ex(Pt)explicit3FEYNMAN}) by replacing the constant surface 
density $\eta$ with radially {\it tapered} density $\eta(r)$ and then 
evaluating the integral,
\BWT
\begin{align}
\label{eq:integration_a_la_Feynman}
\int_z^{\infty} \, \eta(r) \, e^{i(\omega/c) r} \, dr
& \approx
\sum_j \eta(r_j) \, e^{i(\omega/c) r_j} \, \Delta r_j 
\nonumber \\
& =
\eta(z) \, e^{i(\omega/c) z} \, \Delta r 
+
\eta(z+\Delta r) \, e^{i(\omega/c) (z+\Delta r)} \, \Delta r 
+
\eta(z+2\Delta r) \, e^{i(\omega/c) (z+2 \Delta r)} \, \Delta r
+ \dots
\nonumber \\
& =
\eta(z) \, e^{i(\omega/c) z}
\left\{ 
\Delta r
+
\Delta r \, \frac{\eta(z+\Delta r)}{\eta(z)} \, e^{i(\omega/c)\Delta r}  
+
\Delta r \, \frac{\eta(z+2 \Delta r)}{\eta(z)} \, e^{2 i(\omega/c) \Delta r} 
+ \dots
\right\},
\end{align}
\EWT
graphically by adding up small arrows of slowly decreasing length that 
represent the  complex-valued terms in the curly brackets of the Riemann sum above.
By depicting this process in Fig.\ \ref{fig:4} we immediately see that the integral is 
equal to $i(c/\omega)\, \eta(z) \, e^{i(\omega/c) z}$, which corresponds to setting 
the infinite exponent in Eq.\  (\ref{eq:F(30.16)}) to zero, thus recovering 
Eq.\  (\ref{eq:F(30.18)}).

\subsection{Derivation continued}

\begin{figure} [!ht]
\includegraphics[angle=0,width=0.6\linewidth]{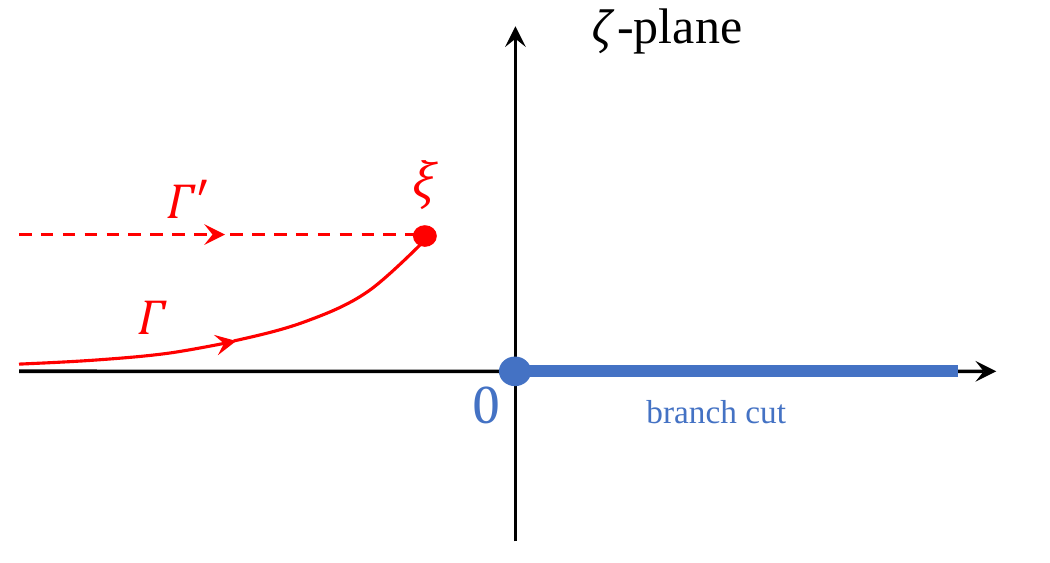}
\caption{ \label{fig:5}  
Sketch of two alternative paths in the complex $\zeta$-plane defining 
the exponential integral function ${\rm Ei}(\xi)$.
}
\end{figure}

We now return to our expression (\ref{eq:Ex(Pt)explicit3}) and regularize 
it \`{a} la Feynman 
by replacing the constant density $\eta$ with the tapered one which we choose to be 
in the form,
\BEq
\eta(r) = \eta e^{-\alpha(\omega/c)r}, \quad \alpha >0,
\EEq
where $\alpha$ is a small positive parameter to be set to zero at the end 
of the calculation.
This gives,
\begin{align}
I(z,a) 
&= 
\int_z^a \, e^{(i-\alpha)(\omega/c) r} 
\Bigg\{
\frac{\omega^2}{c^2} 
- \frac{i\omega}{c} \, \frac{1}{r}
+\left(1+\frac{\omega^2 z^2}{c^2}\right)\frac{1}{r^2}
 + \frac{3i\omega z^2}{c} \frac{1}{r^3} 
-\frac{3z^2}{r^4}
\Bigg\}
\, dr \,,
\end{align}
which allows us to interpret Feynman's regularization procedure 
as Wick rotation in the complex $r$-plane (see also Fig.\ \ref{fig:6} 
and Eq.\ (\ref{eq:Ei_limit_a_infty}) below). 
Next,
\BWT
\begin{align}
I(z,a) 
&=
\frac{\omega}{(i-\alpha)c}
\left(e^{\xi_2} - e^{\xi_1} \right)
- \frac{i\omega}{c} 
\int_{\xi_1}^{\xi_2} 
\frac{e^{\zeta}}{\zeta}\, d \zeta
+
\left(1+\frac{\omega^2 z^2}{c^2}\right)
(i-\alpha)\,\frac{\omega}{c}
\int_{\xi_1}^{\xi_2} 
\frac{e^{\zeta}}{\zeta^2}\, d \zeta 
\nonumber \\
&\quad
+
\frac{3i\omega z^2}{c} 
\left[(i-\alpha)\,\frac{\omega}{c}\right]^2
\int_{\xi_1}^{\xi_2} 
\frac{e^{\zeta}}{\zeta^3}\, d \zeta 
-3z^2 
\left[(i-\alpha)\,\frac{\omega}{c}\right]^3
\int_{\xi_1}^{\xi_2} 
\frac{e^{\zeta}}{\zeta^4}\, d \zeta \,,
\end{align}
\EWT
where
\BEq
\xi_2 = (i-\alpha)(\omega/c)a,
\quad
\xi_1 = (i-\alpha)(\omega/c)z.
\EEq
Taking into account that
\BWT
\begin{align}
\int_{\xi_1}^{\xi_2} \frac{e^{\zeta}}{\zeta^2}\, d \zeta
&=
\int_{\xi_1}^{\xi_2} \frac{e^{\zeta}}{\zeta}\, d \zeta
 - \left. \frac{e^{\zeta}}{\zeta}\right\vert^{\xi_2}_{\xi_1},
\nonumber \\
\int_{\xi_1}^{\xi_2} \frac{e^{\zeta}}{\zeta^3}\, d \zeta
&=
\frac{1}{2}
\left\{
\int_{\xi_1}^{\xi_2} \frac{e^{\zeta}}{\zeta}\, d \zeta
 - 
\left[
 \frac{e^{\zeta}}{\zeta}
+
\frac{e^{\zeta}}{\zeta^2}
\right]^{\xi_2}_{\xi_1}
\right\},
\nonumber \\
\int_{\xi_1}^{\xi_2} \frac{e^{\zeta}}{\zeta^4}\, d \zeta
&=
\frac{1}{2\cdot 3}
\left\{
\int_{\xi_1}^{\xi_2} \frac{e^{\zeta}}{\zeta}\, d \zeta
 - 
\left[
\frac{e^{\zeta}}{\zeta}
+
\frac{e^{\zeta}}{\zeta^2}
+\frac{2 e^{\zeta}}{\zeta^3}\right]^{\xi_2}_{\xi_1}
\right\},
\end{align}
\EWT
and
\begin{align}
\int_{\xi_1}^{\xi_2} \frac{e^{\zeta}}{\zeta}\, d \zeta
&=
{\rm Ei}(\xi_2)-{\rm Ei}(\xi_1),
\end{align}
after some algebra we find,
\BWT
\begin{align}
\label{eq:Ifinal}
I(z,a) 
&=
e^{(i-\alpha) (\omega/c)a} 
\Bigg[
\frac{\omega}{(i-\alpha) c}
-\frac{1}{a}
\left(
1-\frac{z^2}{a^2}+\frac{(\alpha +2 i) \omega z^2}{2 a c}
-\frac{\alpha  (\alpha +i) \omega^2 z^2}{2 c^2}
\right)
\Bigg]
\nonumber \\
&\quad
-\frac{\omega}{2 c} \frac{e^{(i-\alpha) (\omega/c)z}}{(i-\alpha)} 
\left((4+\alpha  (\alpha +i))  
-\alpha  \left(\alpha ^2+1\right) \frac{\omega z}{c}\right)
\nonumber \\
&\quad
-\frac{\alpha  \omega}{c}
 \left(1-\left(\alpha ^2+1\right) \frac{\omega^2 z^2}{2 c^2}\right)
\left({\rm Ei}\left[(i-\alpha)\frac{\omega}{c}a\right] 
- {\rm Ei}\left[(i-\alpha)\frac{\omega}{c}z\right]\right),
\end{align}
\EWT
where ${\rm Ei}(\xi) $ is the well-known exponential integral  
\cite{lebedev1972special},
\BEq
{\rm Ei}(\xi) = \int_{-\infty}^{\xi} \frac{e^{\zeta}}{\zeta}\, d \zeta \,,
\quad
|{\rm arg}(-\xi)|<\pi \,.
\EEq
The restriction on the argument is due to the standard choice of the 
branch cut that makes the integral single-valued (Fig.\ \ref{fig:5}).

\begin{figure} [!ht]
\includegraphics[angle=0,width=0.6\linewidth]{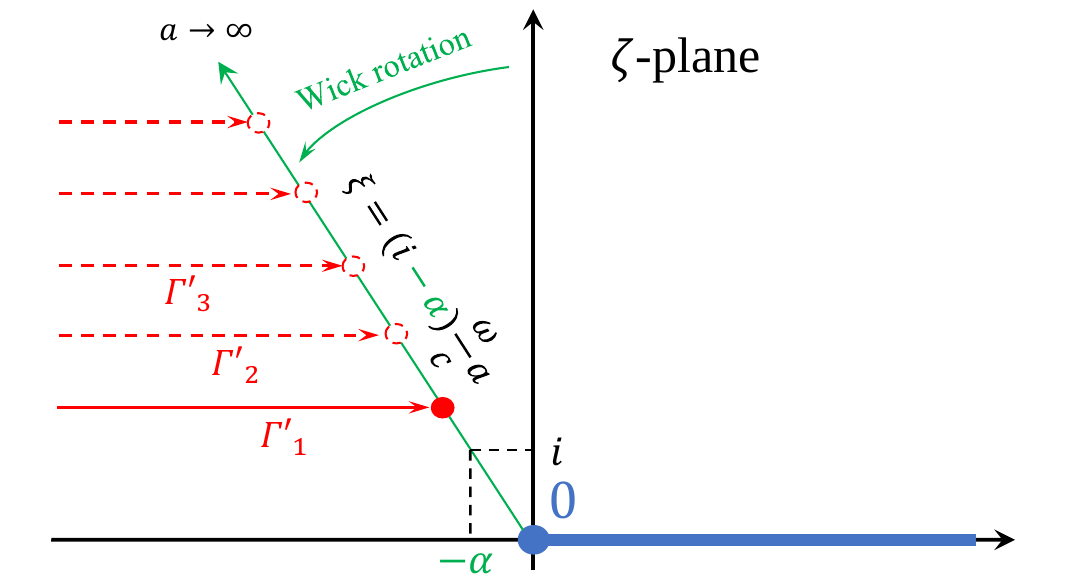}
\caption{ \label{fig:6}  
The limit of ${\rm Ei}(\xi)$ as $a\rightarrow \infty$ along the ray
$\xi = (i-\alpha)\frac{\omega}{c}a$ is zero.
}
\end{figure}

We can view ${\rm Ei}(\xi)$ as accumulation function for
$f(\z)=e^{\z}/\z$ along a path (call it $\Gamma$) running from 
$-\infty$ to $\xi$ in the complex $\z$-plane. The choice of this defining 
path is not unique. One may deform the path arbitrarily as long as it 
originates at the ``left'' infinity and does not cross the branch cut. 
A simple example is provided by the semi-infinite straight segment 
$\Gamma'$ that runs parallel to the real axis and is described by 
$-\infty < {\rm Re}(\z) \leq {\rm Re}(\xi)$, 
${\rm Im}(\z) = {\rm Im}(\xi)$. 
Based on this observation we can see that the limit of ${\rm Ei}(\xi)$ 
as $\xi \rightarrow \infty$ along the ray shown in Fig.\ \ref{fig:6} is 
equal to zero, 
\BEq
\label{eq:Ei_limit_a_infty}
\lim_{a\rightarrow \infty} {\rm Ei}\left[(i-\alpha)\frac{\omega}{c}a\right] =0.
\EEq 
With this result at hand, we are ready to complete 
our derivation (see Eqs.\ (\ref{eq:Ex(Pt)explicit3}) and (\ref{eq:Ifinal})).

\subsection{Infinite plane}

In the case of infinite plane we take the double-limit in (\ref{eq:Ifinal}) 
using (\ref{eq:Ei_limit_a_infty}) by first letting  
$a\rightarrow \infty$ and then $\alpha \rightarrow 0$. 
This immediately gives,
\BEq
I 
=
2i\,\frac{\omega}{c}\, e^{i(\omega/c)z}\,,
\EEq
whose use in (\ref{eq:Ex(Pt)explicit3}) recovers Feynman's formula, 
\BEq
\label{eq:INFINITEplane}
E_x(z,t)
= 
2\pi \, \frac{\eta q}{c}\, i\omega x_0 e^{-i\omega (t-z/c)}
\quad {\rm (infinite\, plane)}.
\EEq

\subsection{Finite disk}

\begin{figure} [!ht]
\includegraphics[angle=0,width=0.55\linewidth]{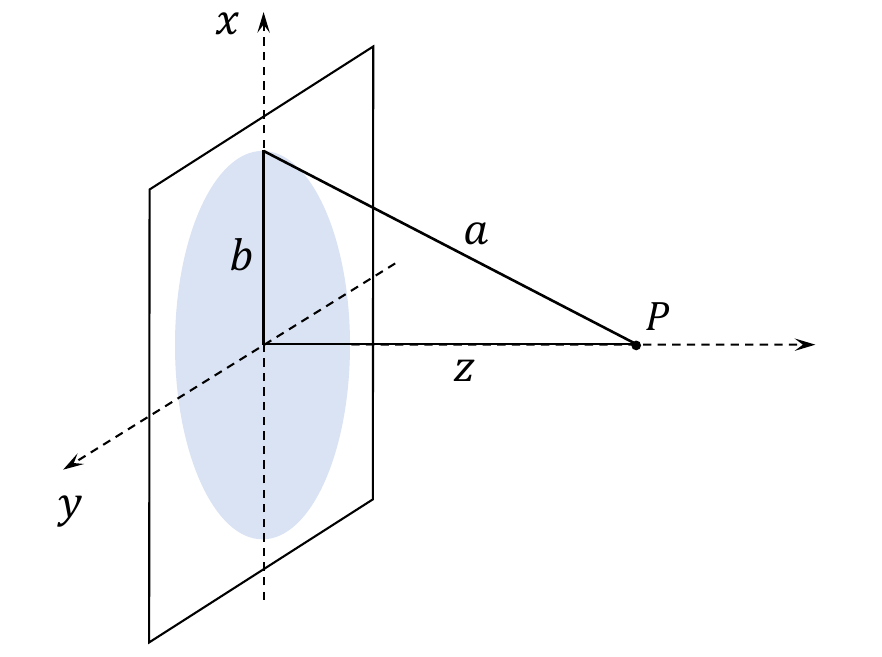}
\caption{ \label{fig:7}  
Radiating disk of radius $b$.
}
\end{figure}

In the case of a finite disk (of radius $b$) we set $\alpha = 0$ in (\ref{eq:Ifinal}), 
keeping $a$ and $z$ finite, as shown in Fig.\ \ref{fig:7}. This gives,
\begin{align}
\label{eq:Edisk}
{E}_x(z,t)
&=
2 \pi \frac{\eta  q }{c} i \omega  x_0 \, e^{-i\omega (t-z/c)} \, {\cal A}(z),
\quad
{\cal A}(z)
=
1
-
\frac{1}{2} 
\left(
1
+
\frac{z^2}{a^2}
-\frac{ic}{{\omega} a}
\frac{b^2}{a^2}
\right)
e^{i (\omega/c)(a-z)},
\quad
a=\sqrt{z^2+b^2}\,,
\end{align}
which is valid everywhere on the symmetry axis. Notice that this expression
vanishes in the limit $z\rightarrow \infty$ (via ${\cal A}\rightarrow 0$).
Also, in the limit $b\rightarrow \infty$ we get
Eq.\ (\ref{eq:INFINITEplane}) for the infinite plane, provided 
we formally set $e^{i\infty}$ in ${\cal A}$ to zero.

\begin{figure} [!ht]
\includegraphics[angle=0,width=0.55\linewidth]{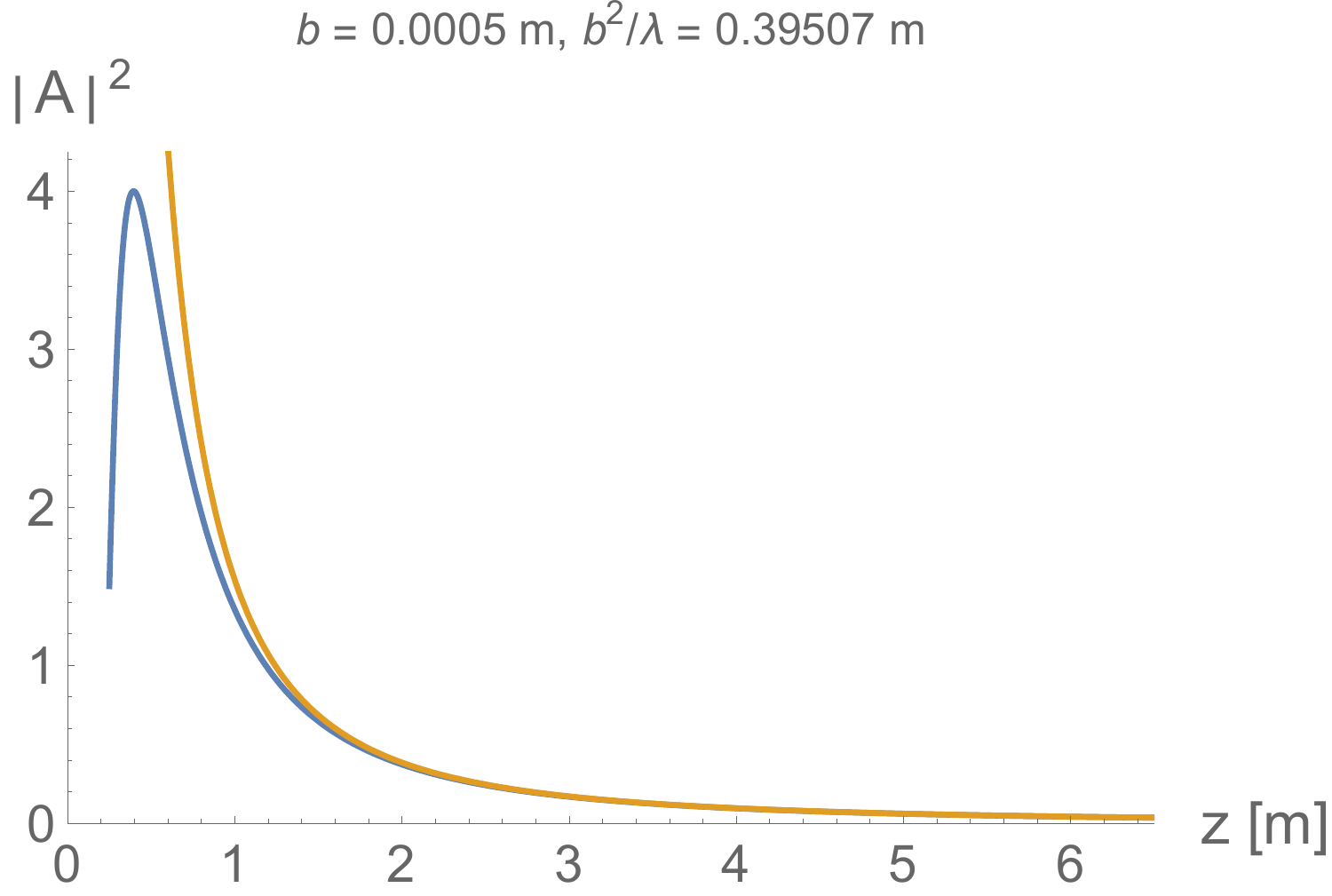}
\vskip15pt
\includegraphics[angle=0,width=0.55\linewidth]{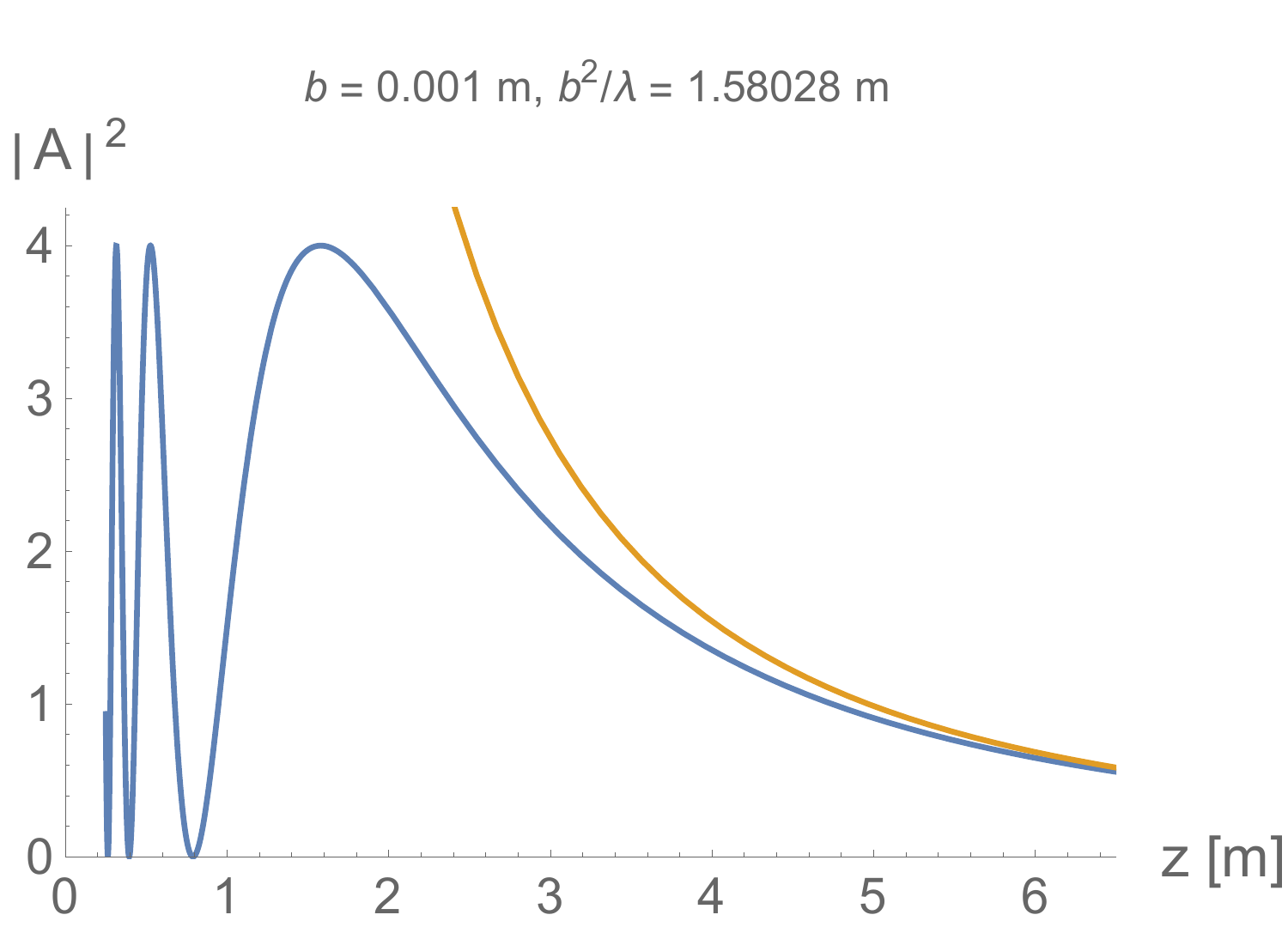}
\vskip15pt
\includegraphics[angle=0,width=0.55\linewidth]{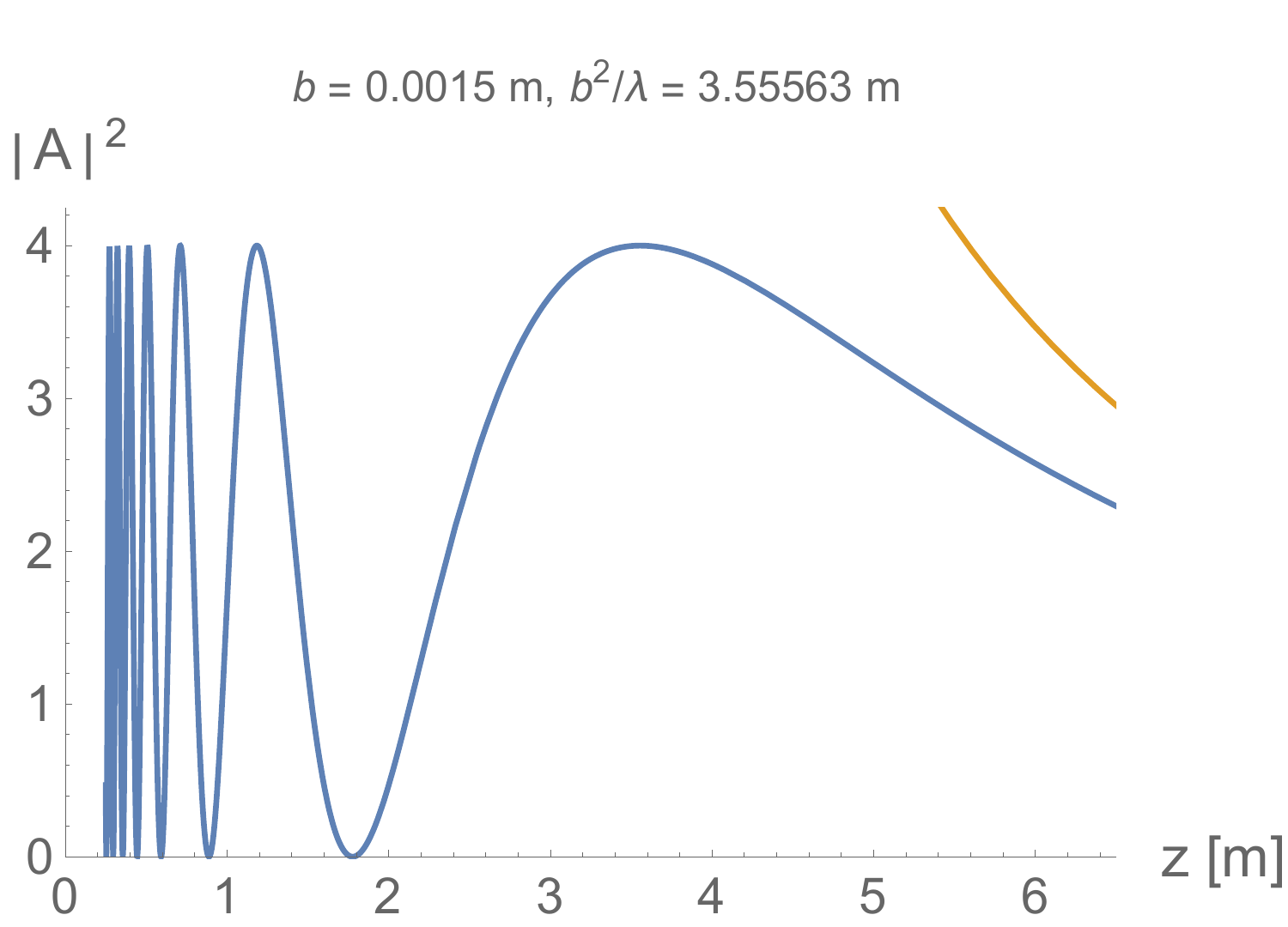}
\caption{ \label{fig:8}  
Blue curves: predicted behavior of the radiation intensity, $|{\cal A}(z)|^2$
({\it on the symmetry axis}, at points $P=(0,0,z)$), 
as the function of the distance, $z$, from a uniform disk of oscillating dipoles, 
as described by Eq.\ (\ref{eq:DISKregime1experimental}).
Here, $\lambda = 632.8$ nm is the radiation wavelength and $b$ is the radius
of the disk.
Orange curves: asymptotic behavior of $|{\cal A}(z)|^2$ at large $z$, as given by 
Eqs.\ (\ref{eq:DISKregime2experimental}) and  (\ref{eq:DISKregime2experimentalPOINTLIKE}).
}
\end{figure}

\begin{figure} [!ht]
\includegraphics[angle=0,width=0.55\linewidth]{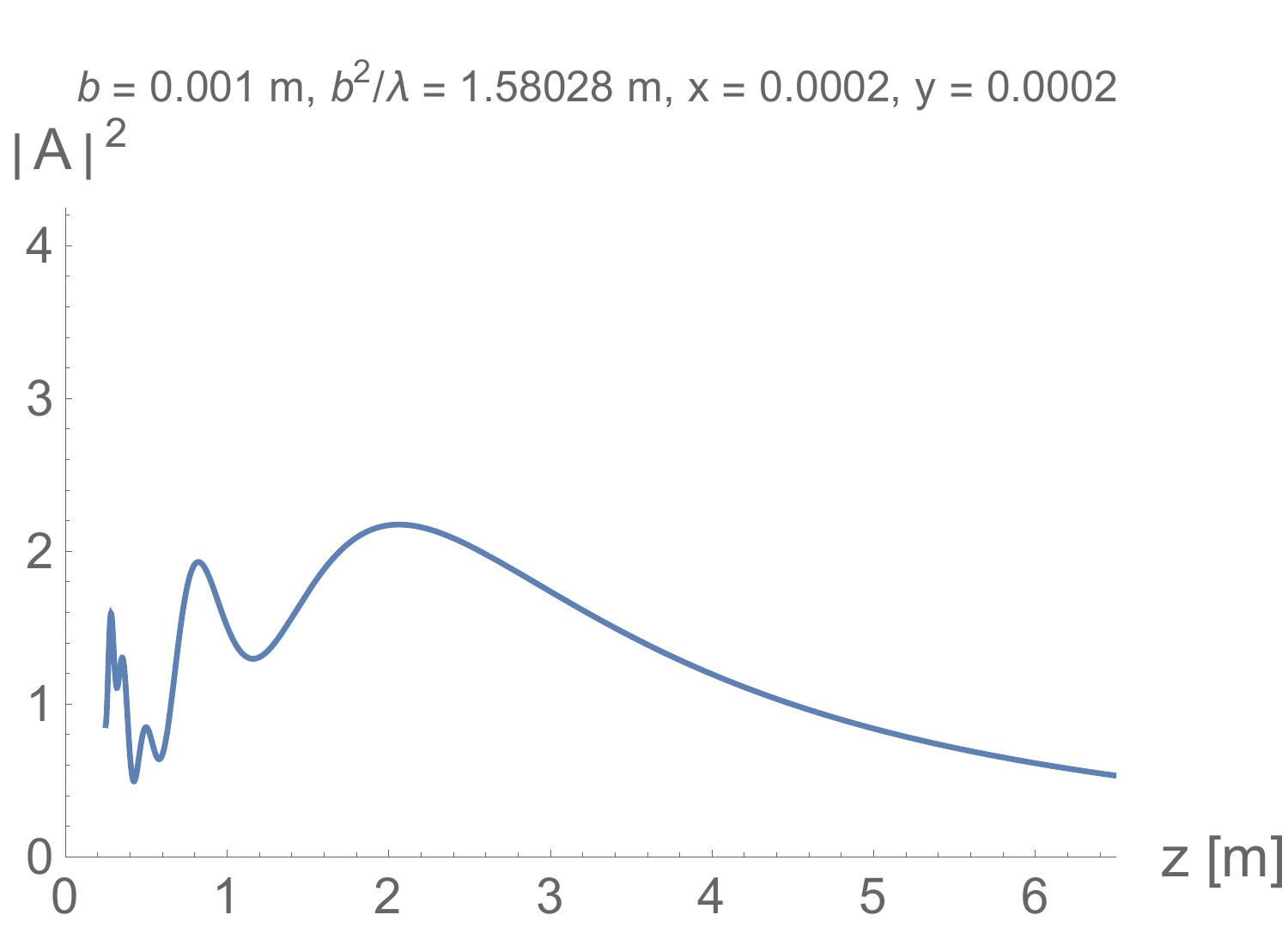}
\vskip15pt
\includegraphics[angle=0,width=0.55\linewidth]{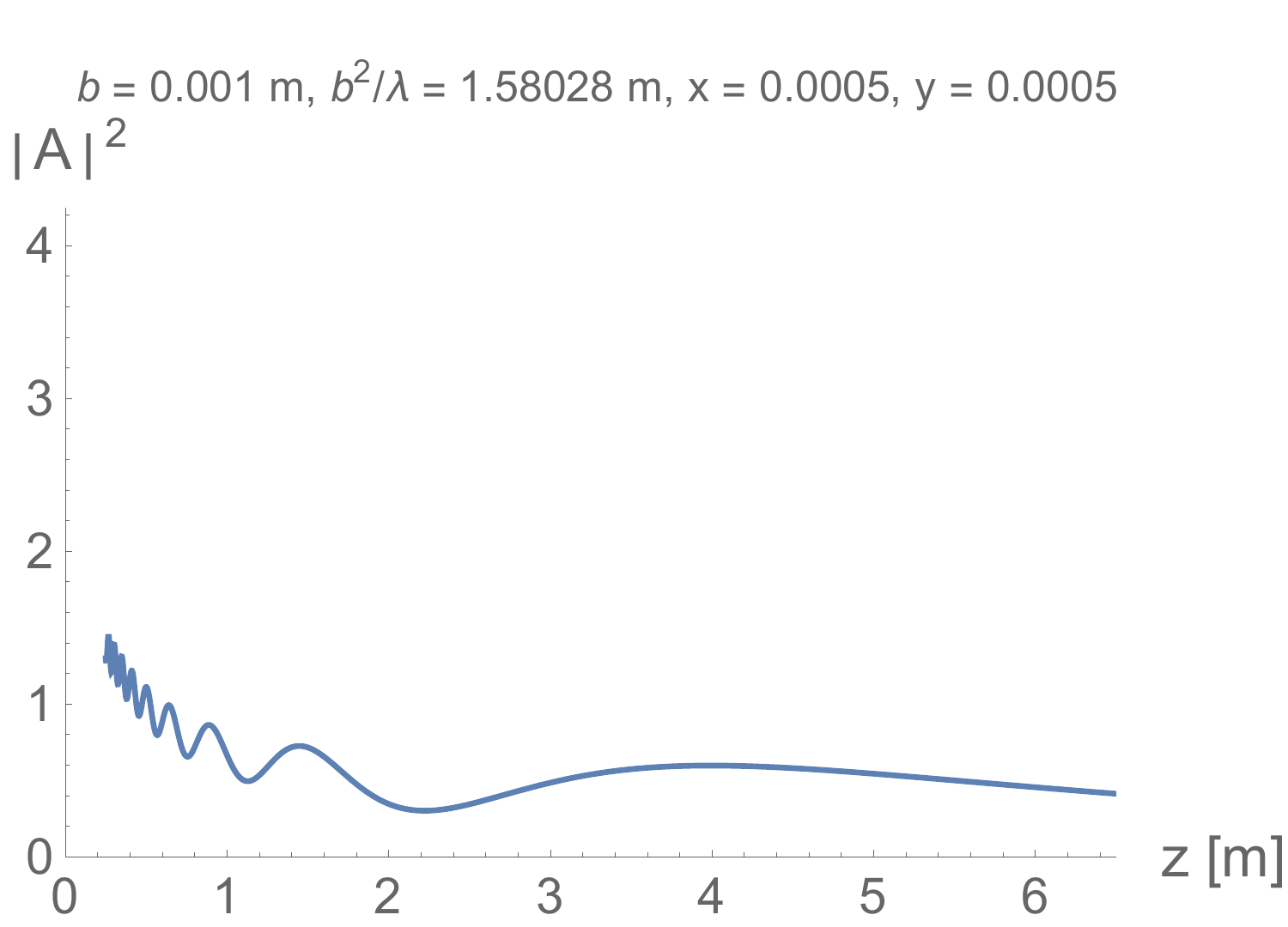}
\vskip15pt
\includegraphics[angle=0,width=0.55\linewidth]{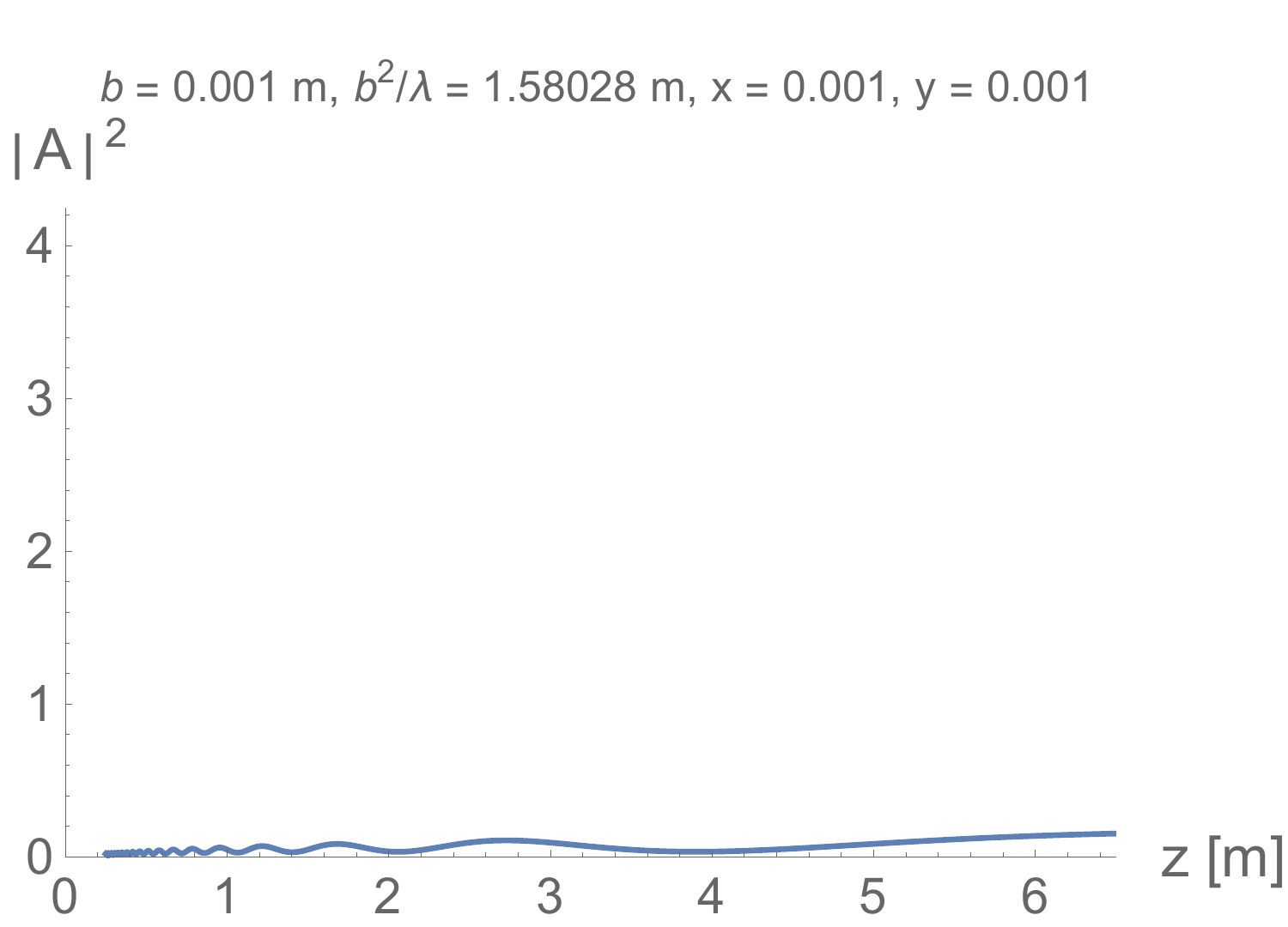}
\caption{ \label{fig:9}  
Blue curves: radiation intensity, $|{\cal A}(z)|^2$ ({\it off the symmetry axis}, 
at points $P=(x,y,z)$), 
as the function of the distance, $z$, from a uniform disk of oscillating dipoles, 
found by numerical integration of Eq.\ (\ref{eq:E(Pt)}).
Here, $\lambda = 632.8$ nm is the radiation wavelength and $b=1$ mm is the radius
of the disk.
}
\end{figure}

For the far field, with
$\lambda/z \ll 1$,  $b/z \ll 1$ (with no additional restriction on $b$),  
using approximation $a\approx z (1 + b^2/(2z^2))$ we get,
\BEq
\label{eq:DISKregime1experimental}
{E}_x(z,t)
=
2\pi \, \frac{\eta q}{c} \,  i \omega x_0 e^{-i\omega (t-z/c)} \, {\cal A}(z) \,,
\quad
{\cal A}(z)
=
1-e^{i \pi b^2/(\lambda z)}  \,,
\EEq
and the corresponding (unnormalized) intensity,
\BEq
\label{eq:DISKregime1experimentalINTENSITY}
|{\cal A}(z)|^2 = 2\left[1-\cos \left(\frac{\pi b^2}{\lambda z}\right)\right],
\EEq
where $\pi b^2$ is the area of the disk.
If, in addition, we impose a stronger condition 
$b^2/(\lambda z)\ll 1$,
we recover the intuitively anticipated result,
\BEq
\label{eq:DISKregime2experimental}
E_x(z,t)
=
2\pi \, \frac{\eta q}{c} \,  i \omega x_0 e^{-i\omega (t-z/c)} \, {\cal A}(z) \,,
\quad
{\cal A}(z)
=
-i \frac{\pi b^2}{\lambda z}  \,,
\quad
|{\cal A}(z)|^2 = \left(\frac{\pi b^2}{\lambda z}\right)^2,
\EEq
or,
\BEq
\label{eq:DISKregime2experimentalPOINTLIKE}
E_x(z,t)
= 
\frac{Q}{c^2 z}\, \omega^2 x_0 \, e^{-i\omega (t-z/c)}\,,
\quad
Q=\pi b^2 \eta q\,,
\EEq
for a ``point-like'' disk of total charge $Q$ viewed from a very large distance.
The predicted intensity of disk's field, $|{\cal A}(z)|^2$, on the symmetry axis,
as the function of the distance, $z$, in a realistic experimental regime described by 
Eq.\ (\ref{eq:DISKregime1experimental}) is shown in Fig.\ \ref{fig:8}.
[For comparison, the intensity {\it off the symmetry axis} is plotted in Fig.\ \ref{fig:9}.]

Before concluding this section, let us make a mental note that the expression for the field 
given by Feynman's formula is proportional to the retarded {\it velocity} 
of the oscillating charges, or, equivalently, to the corresponding retarded 
electric {\it current}. The appearance of the current is not too surprising 
and may be traced to a more traditional approach to scattering on two-dimensional structures, 
as we briefly outline in Appendix \ref{appendix:1}. When we come to discuss electrodynamics of 
chiral media, this result, properly generalized, will play a 
central role in connecting our theory to the previously published work on 
metamaterial nanoplasmonics. Such connection is possible because within 
the Born-Kuhn model of optical activity the very definition of optical 
rotatory dispersion is given in terms of the bound {\it currents} induced 
in the material by the incident electromagnetic wave (see, e.\ g., discussion in  \cite{davis2019microscopic}). 

\clearpage

\section{Applications in metamaterial nanoplasmonics}

\subsection{Laser beam striking a sheet of aligned plasmonic nanorods}

\begin{figure} [!ht]
\includegraphics[angle=0,width=0.85\linewidth]{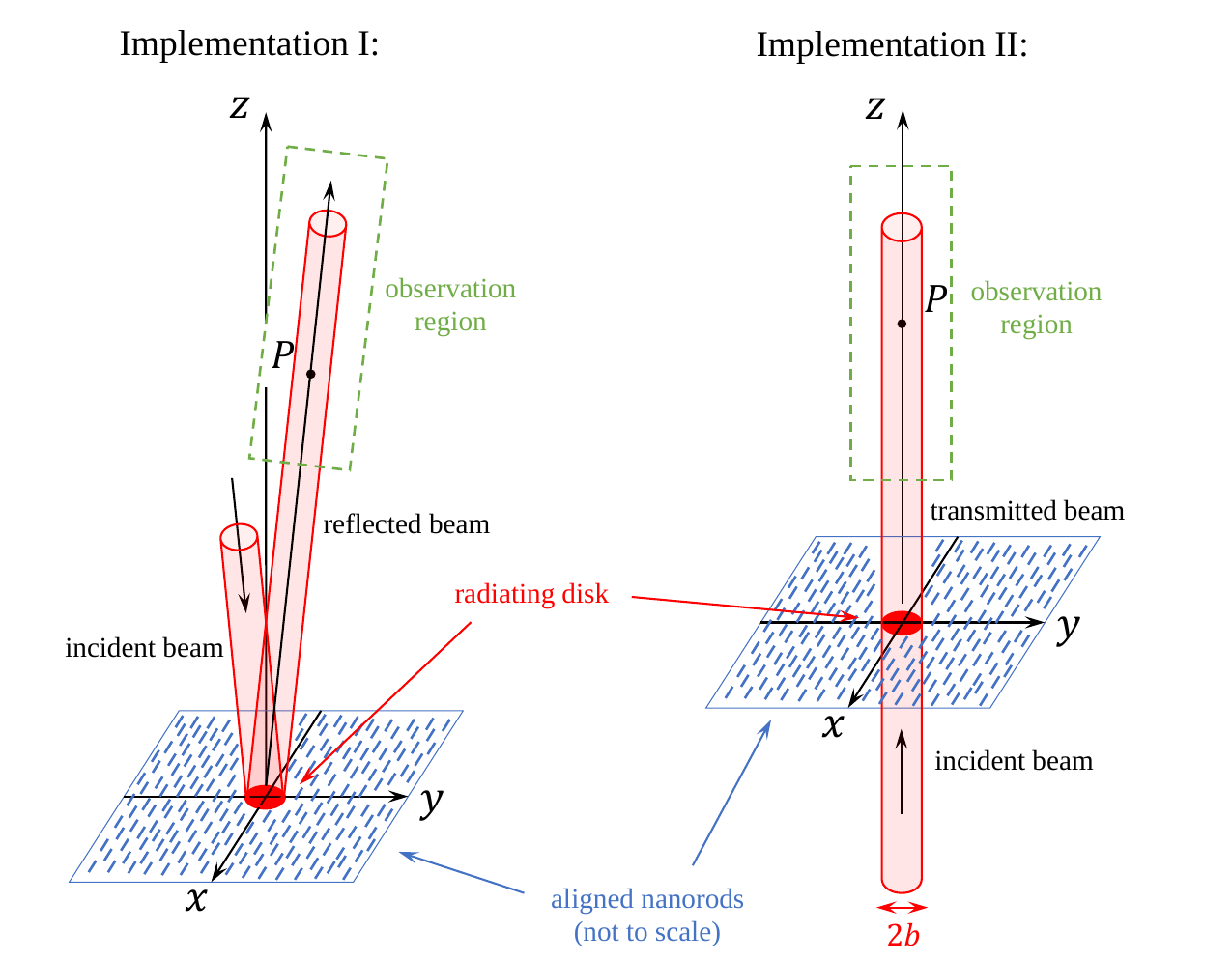}
\caption{ \label{fig:10}  
Schematics of the experiment in which radiating disk of oscillating dipoles is effectively 
generated by a laser beam that either (I) is scattered (at nearly normal incidence) 
by, or (II) passes through, a 
two-dimensional sheet of aligned plasmonic nanorods.}
\end{figure}

Assume now that the radiating disk of finite radius $b$ 
is artificially induced by a uniform laser beam 
of wavelength $\lambda = 632.8$ nm 
(He-Ne laser, red) and cross-sectional radius $b = 1$ mm striking infinitesimally thin 
sheet of dipoles, such as a layer of plasmonic nanorods, as depicted 
in Fig.\ \ref{fig:10}. Assume also that the observation point $P$ is chosen, say, 
anywhere between 0.25 to 6.5 meters from the sheet. In that case,
\BEq
\omega b/c =2\pi b/\lambda =2\pi \times 1580\,,
\EEq
and
\BEq
0.00015 < b/z < 0.004,
\quad
0.24 < b^2/(\lambda z) < 6.3,
\quad
b^2/\lambda = 1.58 {\rm \, m}\,,
\EEq
so we are in the experimental situation described by Eq.\ (\ref{eq:DISKregime1experimental}). 
For Implementation I, in which the field measurements are made in the reflected beam, 
the predicted intensity, $|{\cal A}(z)|^2$, as the function of 
the distance, $z$, is given in Fig.\ \ref{fig:8} (provided we work at nearly normal incidence).
Despite the high sensitivity of the intensity pattern to the values of the beam's diameter,
the proposed setup seems to be within an easily realizable range of experimental 
parameters, which should make the observation of this interference effect relatively 
straightforward. [Here we visualize a simplified scenario in which the nanorods are 
not too densely packed, potentially missing on some interesting contributions coming 
from various near field interaction effects 
(see, e.\ g., \cite{lagarkov1996electromagnetic,ivanov2012plasmonic};
also Appendix \ref{appendix:1}).]

For Implementation II, in which the field measurements are made in 
the transmitted beam, the net field at $P$ is the sum of the incoming wave
$({\cal E}_{x},{\cal E}_{y})\, e^{-i\omega (t-z/c)}$ and the radiation field of the oscillating disk. 
The equation of motion for a driven dipole in this case is
\begin{align}
\ddot{x}+\gamma \dot{x}+\omega_{0}^2 x 
&= (q/m) {\cal E}_{x}e^{-i\omega t},
\end{align} 
where $\omega_{0}$ is the natural frequency of the equivalent Drude-Lorentz 
oscillator, $\gamma$ is the damping coefficient, and $q/m$ is the charge-to-mass ratio. 
Denoting
\BEq
\Omega^2 \equiv \omega_{0}^2 - \omega^2 -i\gamma \omega ,
\EEq
we get the steady state solution,
\BEq
\label{eq:xyPositions}
    x(t) = x_0  e^{-i\omega t},
\quad
x_0
=
\frac{q}{m \Omega^2}{\cal E}_{x},
\EEq
which, on the basis of (\ref{eq:DISKregime1experimental}), gives the total 
transmitted field,
\begin{align}
\label{eq:ExIMPLEMENTATIONII}
E_x(z,t)
=
{\cal T}(z)  \, {\cal E}_x \, e^{-i\omega (t-z/c)},
\quad
E_y(z,t)
=
{\cal E}_y \, e^{-i\omega (t-z/c)},
\quad
{\cal T}(z)
\equiv 
1 +i \, \frac{\omega D}{2c} \frac{\omega_{{\rm p}1}^2}{\Omega^2}
\, {\cal A}(z),
\end{align}
where ${\cal A} = 1$ for a wide beam and 
${\cal A}(z) = 1-e^{i \pi b^2/(\lambda z)}$ for a narrow beam of 
cross-sectional radius $b$. In the above, we formally 
introduced the plasma frequency (with Coulomb's constant restored) 
for a {\it single} sheet of nanorods, hence subscript 1 ({\it cf}.\ (\ref{eq:wp})),
\BEq
\label{eq:wp1}
\omega_{{\rm p}1}^2 \equiv \frac{\eta q^2}{\varepsilon_0 m D}\,,
\EEq
where $D$ is the thickness of the metamaterial plate (practically, nanorod's
cross-sectional diameter) and $\varepsilon_0$ is the electric constant. 
Notice that a single layer of aligned nanorods is functionally equivalent to a 
birefringent plate. Also notice that the transmission amplitude can be written in a slightly different form,
\BEq
{\cal T}(z)
= 
1 +i \, \frac{\omega  \omega_{{\rm p}1}'}{\Omega^2}
\, {\cal A}(z),
\EEq
in terms of a different characteristic frequency (primed plasma frequency),
\BEq
\label{eq:wp1PRIMED}
\omega_{{\rm p}1}' \equiv \frac{\eta q^2}{2\varepsilon_0 m c}\,,
\EEq
which may be useful in some contexts, though in what follows we will adhere to
traditional definition.

\begin{figure} 
\includegraphics[angle=0,width=0.6\linewidth]{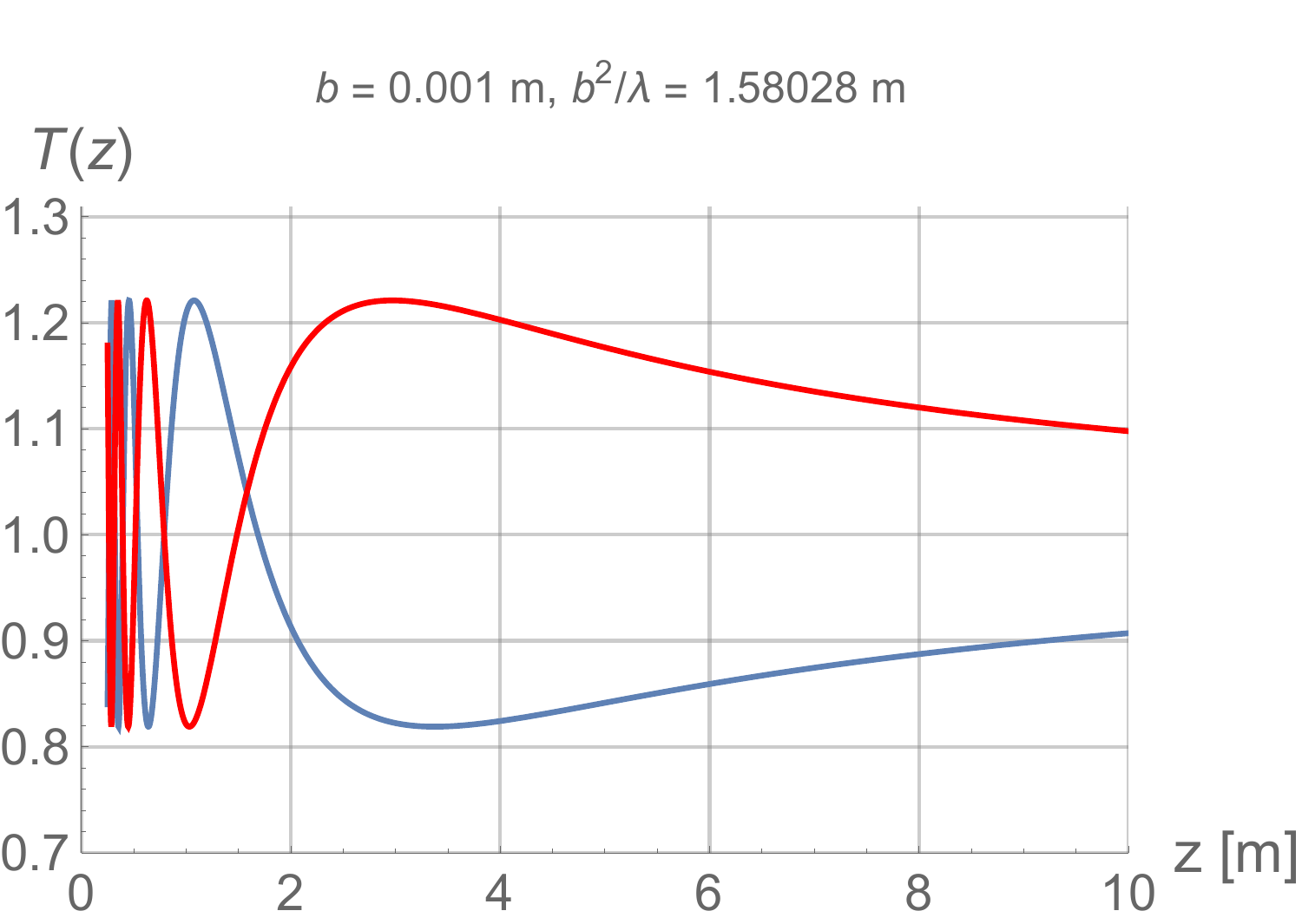}
\caption{ \label{fig:11}  
Predicted transmitted intensity, $T(z)=|{\cal T}(z)|^2$ (unnormalized), 
as the function of the distance, $z$, Eq.\ (\ref{eq:DISKregime3experimentalINTENSITY}),
for an $x$-polarized laser beam passing through a sheet of aligned plasmonic nanorods
(Implementation II). 
Beam parameters: $\lambda = 632.8$ nm, $b=1$ mm, with 
arbitrarily chosen $(n_{1}-1)\omega D/c=0.1$ (red curve, 
corresponding to $\omega_0 > \omega$) and 
$(n_{1}-1)\omega D/c=-0.1$ (blue curve, 
corresponding to $\omega_0 < \omega $).
}
\end{figure}

The ``usual'' theory based on macroscopic electrodynamics 
is recovered 
by setting ${\cal A}=1$ (wide beam) and introducing the quantity 
which we may call the index of refraction (for lack of a better word), 
\BEq
n_{1} \equiv 1 + \frac{ \omega_{{\rm p}1}^2}{2\Omega^2}\,,
\EEq
which for $\omega$ far from resonant $\omega_0$ corresponds approximately 
to the permittivity of the Drude-Lorentz model, 
$n_{1}^2 \approx \varepsilon_{\rm DL} = 1 + \omega_{{\rm p}1}^2/\Omega^2$. 
In the absence of losses ($\gamma = 0$) and for $\omega \ll \omega_0$, the 
index of refraction is real and positive, so for an $x$-polarized wave we get,
\begin{align}
E_x(z,t)
=
\left( 
1 +\frac{i\omega (n_{1}-1) D}{c} \right)  \, {\cal E}_x \, e^{-i\omega (t-z/c)}
\approx
\, {\cal E}_x 
\, \exp\left\{-i\omega \left[t-\frac{z}{c}-\left(\frac{D}{c/n_{1}} 
- \frac{D}{c}\right)\right]\right\}\, ,
\end{align}
which shows an increase in the optical path-length due to an effective 
reduction in phase velocity
(from the original $c$ down to $c/n_{1}$) when propagating through the plate, 
in agreement with Feynman's reasoning (\cite{feynman2006feynman}, Ch.\ 31).
In the case of a laser beam, using ${\cal A}(z) = 1-e^{i \pi b^2/(\lambda z)}$ 
and assuming $\gamma = 0$ in Eq.\ (\ref{eq:ExIMPLEMENTATIONII}), we 
get for transmitted radiation intensity,
\BEq
\label{eq:DISKregime3experimentalINTENSITY}
|{\cal T}(z)|^2 
= 
1
+
2 (n_{1}-1)\frac{\omega D}{c} \sin \left(\frac{\pi b^2}{\lambda z}\right)
+
2 \left((n_{1}-1)\frac{\omega D}{c}\right)^2 
\left[1-\cos \left(\frac{\pi b^2}{\lambda z}\right)\right],
\EEq
which is depicted in Fig.\ \ref{fig:11}.

\subsection{Laser beam striking a chiral plate}

\begin{figure} 
\includegraphics[angle=0,width=0.75\linewidth]{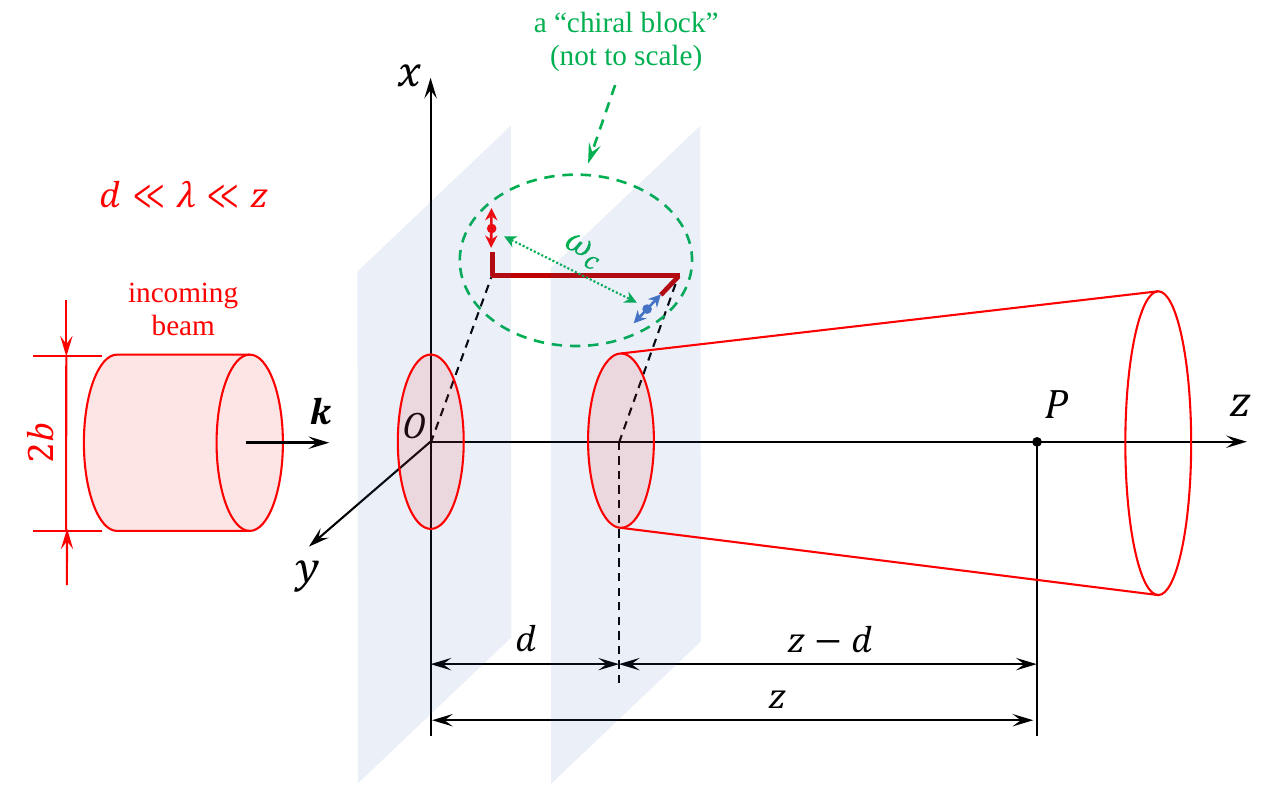}
\caption{ \label{fig:12}  
Schematic representation of the fundamental structural element (the "chiral block")
of a gyrotropic metamaterial in accordance with the two-dimensional Born-Kuhn 
oscillator model.
One possible implementation of the model consists of a single layer of such blocks 
which, in turn, can be visualized as being made of two planes of pair-wise coupled 
harmonically oscillating charges.
}
\end{figure}

The two-dimensional Born-Kuhn (BK) oscillator model 
\cite{kuhn1930physical, born1935theory} 
(also \cite{svirko2000polarization, schaferling2017chiral})
represents a chiral block (the fundamental structural element of a chiral medium) 
by two coupled charged harmonic oscillators, one at $z=0$ and the other at 
$z=d$, displaced as depicted in Fig.\ \ref{fig:12} and subjected to a plane 
electromagnetic wave of frequency $\omega$ and wave number $k=\omega/c$ 
propagating with the limiting speed $c$  along the $z$-axis. 
The charges are restricted to move in the $x$ and $y$ directions, respectively, in 
which case their equations of motion are
given by
\begin{align}
\ddot{x}+\gamma \dot{x}+\omega_{0}^2 x 
 +\omega_{\rm c}^2 y &= (q/m) {\cal E}_{x}e^{-i\omega t},
\\
\ddot{y}+\gamma \dot{y}+\omega_{0}^2 y 
+\omega_{\rm c}^2 x &= (q/m) {\cal E}_{y}e^{-i(\omega t - kd)},
\end{align}
where $\omega_{\rm c}$ is the coupling frequency and $({\cal E}_{x}, {\cal E}_{y})$ 
are the amplitudes of the $x$ and $y$ components of the electric field in the incoming 
wave.
Using the steady state {\it Ansatz},
\BEq
\label{eq:xyPositionsChiral}
    x(t) = x_0  e^{-i\omega t},
\quad
    y(t) = y_0 e^{-i(\omega t - kd)},
\EEq
the equations of motion can be written in matrix form,
\begin{align}
\label{eq:equationsOfMotionGeneral}
  \begin{pmatrix}
    \Omega^2 &             \omega_{\rm c}^2 e^{ikd}\\
                \omega_{\rm c}^2 e^{-ikd}& \Omega^2  
  \end{pmatrix}
  \begin{pmatrix}
    x_0 \\
    y_0 
  \end{pmatrix}
=
\frac{q}{m}
  \begin{pmatrix}
    {\cal E}_{x}\\
    {\cal E}_{y}
  \end{pmatrix},
\end{align}
with the solution,
\begin{align}
\label{eq:xySolutionsChiral}
x_0
=
\frac{q \, \Omega^2}{m(\Omega^4-\omega_{\rm c}^4)}
\left(
{\cal E}_{x}-\frac{\omega_{\rm c}^2}{\Omega^2}e^{ikd}{\cal E}_{y}
\right),
\quad
y_0
=
\frac{q \, \Omega^2}{m(\Omega^4-\omega_{\rm c}^4)}
\left(
{\cal E}_{y}-\frac{\omega_{\rm c}^2}{\Omega^2}e^{-ikd}{\cal E}_{x}
\right).
\end{align}
The net field at the observation point is the sum of the incoming field 
and the field of all radiating charges, {\it viewed as two planes of pairwise coupled 
BK oscillators}, which gives, using our 
generalization of Feynman's formula (\ref{eq:DISKregime1experimental}),
\begin{align}
\label{eq:ExTransmittedChiral}
E_x(z,t)
&=
\left\{
{\cal E}_x 
+ 
\frac{i\omega d}{2c}
\frac{ \omega_{\rm p}^2 \Omega^2}{\Omega^4-\omega_{\rm c}^4}
\left(
{\cal E}_{x}-\frac{\omega_{\rm c}^2}{\Omega^2}e^{ikd}{\cal E}_{y}
\right)
 {\cal A}(z)
\right\}  e^{-i\omega (t-z/c)},
\\
\label{eq:EyTransmittedChiral}
E_y(z,t)
&=
\left\{
{\cal E}_y 
+ 
i\,\frac{\omega d}{2c}\,
\frac{\omega_{\rm p}^2\Omega^2}{\Omega^4-\omega_{\rm c}^4}
\left(
{\cal E}_{y}-\frac{\omega_{\rm c}^2}{\Omega^2}e^{-ikd}{\cal E}_{x}
\right)
 {\cal A}(z-d)
\right\}  e^{-i\omega (t-z/c)},
\end{align}
where ${\cal A} = 1$ for a wide and  
${\cal A}(z) = 1-e^{i \pi b^2/(\lambda z)}$ for a narrow beam, respectively,
with the plasma frequency now defined by ({\it cf}.\ (\ref{eq:wp1})),
\BEq
\label{eq:wp}
\omega_{{\rm p}}^2 \equiv \frac{\eta q^2}{\varepsilon_0 m d}\,.
\EEq

\begin{figure}  [!ht]
\includegraphics[angle=0,width=0.5\linewidth]{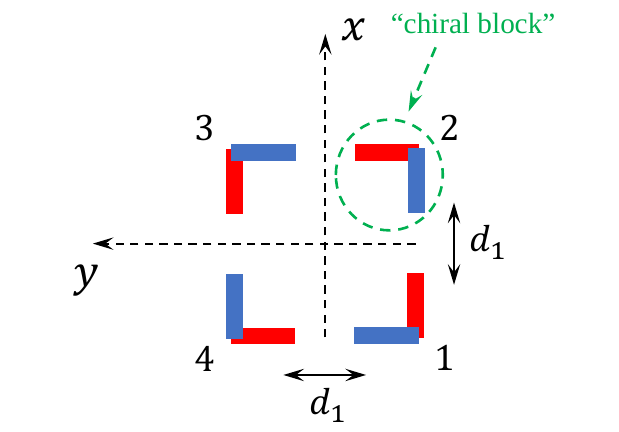}
\caption{ \label{fig:13}  
Top view of a metamaterial supercell consisting of four Born-Kuhn chiral blocks 
arranged in a $C_{4}$-symmetric configuration.
}
\end{figure}

In actual experiments, a BK chiral block is made of 
two identical corner-stacked ``vertically'' displaced plasmonic nanorods that are 
coupled either capacitively or conductively through their mutual near-field 
interaction \cite{yin2013interpreting}
(for an earlier theoretical proposal involving inductive coupling see 
\cite{svirko2001layered}). 
In addition,
in order to avoid anisotropy effects that lead to polarization conversion, the nanorods 
are arranged in $C_{4}$-symmetric configurations (with four nanorod pairs at a time), 
forming square-shaped ``supercells'' out of which metamaterial plates are constructed
(Fig.\ \ref{fig:13}; for details see \cite{yin2013interpreting}).
The corresponding equations of motion for pairs 1 and 2 are then
(notice the physically motivated minus signs in front of some 
of the coupling terms),
\begin{align}
\ddot{x}_1+\gamma \dot{x}_1+\omega_{0}^2 x_1 
 +\omega_{\rm c}^2 y_1 &= (q/m) {\cal E}_{x}e^{-i\omega t},
\\
\ddot{y}_1+\gamma \dot{y}_1+\omega_{0}^2 y_1 
+\omega_{\rm c}^2 x_1 &= (q/m) {\cal E}_{y}e^{-i(\omega t - kd)},
\\
\ddot{x}_2+\gamma \dot{x}_2+\omega_{0}^2 x_2 
 -\omega_{\rm c}^2 y_2 &= (q/m) {\cal E}_{x}e^{-i(\omega t - kd)},
\\
\ddot{y}_2+\gamma \dot{y}_2+\omega_{0}^2 y_2 
-\omega_{\rm c}^2 x_2 &= (q/m) {\cal E}_{y}e^{-i\omega t},
\end{align}
with the solution,
\begin{align}
x_{01}
=
\frac{q \, \Omega^2}{m(\Omega^4-\omega_{\rm c}^4)}
\left(
{\cal E}_{x}-\frac{\omega_{\rm c}^2}{\Omega^2}e^{ikd}{\cal E}_{y}
\right),
\quad
y_{01}
=
\frac{q \, \Omega^2}{m(\Omega^4-\omega_{\rm c}^4)}
\left(
{\cal E}_{y}-\frac{\omega_{\rm c}^2}{\Omega^2}e^{-ikd}{\cal E}_{x}
\right),
\\
x_{02}
=
\frac{q \, \Omega^2}{m(\Omega^4-\omega_{\rm c}^4)}
\left(
{\cal E}_{x}+\frac{\omega_{\rm c}^2}{\Omega^2}e^{-ikd}{\cal E}_{y}
\right),
\quad
y_{02}
=
\frac{q \, \Omega^2}{m(\Omega^4-\omega_{\rm c}^4)}
\left(
{\cal E}_{y}+\frac{\omega_{\rm c}^2}{\Omega^2}e^{ikd}{\cal E}_{x}
\right),
\end{align}
and similarly for pairs 3 and 4. Taking into account that $d\ll z$, so that  
${\cal A}(z-d)\approx {\cal A}(z)$, we get the total transmitted field,
\begin{align}
E_x(z,t)
&=
\left\{
\left(
1
+ 
i\,\frac{\omega d}{2c}\,
\frac{4 \omega_{\rm p}^2 \Omega^2}{\Omega^4-\omega_{\rm c}^4} \, {\cal A}(z)
\right)
{\cal E}_{x}
+
\frac{\omega d}{2c}\,
\frac{4 \omega_{\rm p}^2 \omega_{\rm c}^2}{\Omega^4-\omega_{\rm c}^4} \,
 \sin(kd) \, {\cal A}(z) \, {\cal E}_{y}
\right\}  e^{-i\omega (t-z/c)},
\\
E_y(z,t)
&=
\left\{
\left(
1
+ 
i\,\frac{\omega d}{2c}\,
\frac{4 \omega_{\rm p}^2 \Omega^2}{\Omega^4-\omega_{\rm c}^4} \, {\cal A}(z)
\right)
{\cal E}_{y}
-
\frac{\omega d}{2c}\,
\frac{4 \omega_{\rm p}^2 \omega_{\rm c}^2}{\Omega^4-\omega_{\rm c}^4} \,
 \sin(kd) \, {\cal A}(z) \, {\cal E}_{x}
\right\}  e^{-i\omega (t-z/c)}.
\end{align}

Because the BK model ignores any coupling between different chiral blocks, to 
make it physically reasonable we have to assume that the separation distance 
$d$ between nanorods in a given BK pair is much smaller than the distance 
$d_1$ between nearest 
such pairs (Fig.\ \ref{fig:13}). Therefore, in the context of Feynman's approach, 
the distance $d_1$ should be viewed as the effective size of individual dipoles 
that appear in the derivation of his formula. Since that derivation was performed 
under the assumption of {\it point dipoles}, we have to assume that 
$d_1 \ll \lambda$ and, thus, $d \ll \lambda $, or $kd \ll 1$. Correspondingly, 
from the point of view of a single chiral block, we are in the long-wavelength 
(or, quasi-static) limit, in which case we may set
\BEq
\sin (kd) \approx kd,
\EEq
and get (our main result),
\begin{align}
\label{eq:ExBK1}
E_x(z,t)
&=
\left\{
\left(
1
+ 
i\,\frac{\omega \chi d}{2c}\, {\cal A}(z)
\right)
{\cal E}_{x}
+
\frac{\omega^2 \Gamma d}{2c^2}\, {\cal A}(z) \, {\cal E}_{y}
\right\}  e^{-i\omega (t-z/c)},
\\
\label{eq:ExBK2}
E_y(z,t)
&=
\left\{
-
\frac{\omega^2 \Gamma d}{2c^2}\, {\cal A}(z) \, {\cal E}_{x}
+
\left(
1
+ 
i\,\frac{\omega \chi d}{2c}\, {\cal A}(z)
\right)
{\cal E}_{y}
\right\}  e^{-i\omega (t-z/c)},
\end{align}
where we formally introduced susceptibility and the nonlocality (or, gyration) 
parameter,
\BEq
\chi
\equiv
\frac{4 \omega_{\rm p}^2 \Omega^2}{\Omega^4-\omega_{\rm c}^4} ,
\quad
\Gamma 
\equiv
\frac{4 \omega_{\rm p}^2 \omega_{\rm c}^2 d}{\Omega^4-\omega_{\rm c}^4},
\EEq
to make connection with their counterparts in the Born-Kuhn model 
\cite{svirko2000polarization, yin2013interpreting} 
(the factors of 4 are due to the number of chiral blocks in the supercell).
Diagonalization of the system (\ref{eq:ExBK1}), (\ref{eq:ExBK2}) immediately 
gives the two eigenmodes (the left (LCP) and the right (RCP) circularly polarized 
electromagnetic waves) and their respective eigenvalues, 
${\cal T}_{{\rm L},{\rm R}}$,
\BEq
\label{eq:Tlaser}
\begin{pmatrix}
{\cal E}_{x} \\
{\cal E}_{y}
\end{pmatrix}_{{\rm L},{\rm R}}
=
\frac{{\cal E}_{0} }{\sqrt{2}}
\begin{pmatrix}
1 \\
\pm i
\end{pmatrix},
\quad
{\cal T}_{{\rm L},{\rm R}}(z)
=
1
+ 
i \, \frac{\omega d}{2c}\left(
 \chi 
\pm
\frac{\omega \Gamma }{c} 
\right) {\cal A}(z),
\EEq
where ${\cal E}_{0}$ is the amplitude of the incoming wave.
When these modes pass through a BK plate, their polarization stays the same, 
while the amplitude is multiplied by the corresponding eigenvalue 
({\it cf}.\ Eq.\ (\ref{eq:ExIMPLEMENTATIONII})), which therefore determines 
the transmitted intensity (as a function of the distance), 
$|{\cal T}_{{\rm L},{\rm R}}(z)|^2$.

\begin{figure} [!ht]
\includegraphics[angle=0,width=0.6\linewidth]{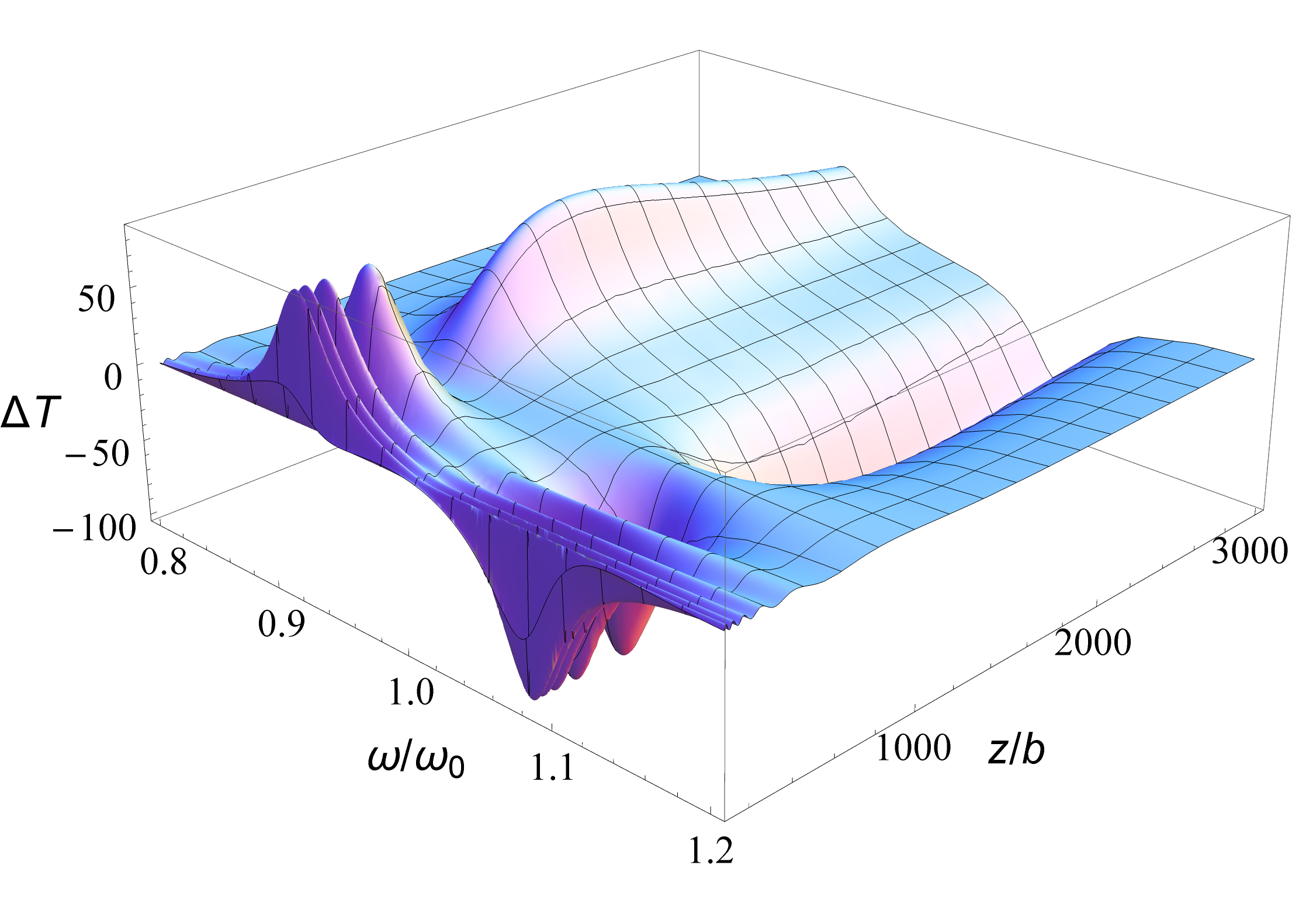}
\caption{ \label{fig:14}  
Transmitted differential intensity, $\Delta T(z)$ (unnormalized), 
as predicted by Eqs.\ (\ref{eq:Tlaser}), (\ref{eq:deltaTlaser}),
for a laser beam of cross-sectional radius $b$, with 
${\cal A}(z) = 1-e^{i \pi b^2/(\lambda z)}$, passing through a 
$C_4$-symmetric BK plate. Here, $b/d=100000$, $kd=0.1$, 
$\omega_{\rm p}/\omega_0=2$,
$\gamma/\omega_0=0.05$, $\omega_{\rm c}/\omega_0=0.4$.
}
\end{figure}

Introducing two ``brightness'' amplitudes,
\BEq
{\cal B}_{{\rm L},{\rm R}}(z)
=
i \, \frac{\omega d}{2c}\left(
 \chi 
\pm
\frac{\omega \Gamma }{c} 
\right) {\cal A}(z),
\EEq
which characterize contributions to the respective transmitted modes due 
to the radiating (portion of the) BK plate, we can define two additional 
experimentally relevant quantities:
differential brightness (which generalizes to narrow beams the concept of 
optical rotatory dispersion (ORD), 
{\it cf}.\ Eq.\ (S20.2) in \cite{davis2019microscopic}),
\BEq
{\rm ORD} 
\sim
\Delta B(z)
\equiv 
|{\cal B}_{{\rm L}}(z)|^2 - |{\cal B}_{{\rm R}}(z)|^2
=
\frac{\omega^2 d^2}{c^2} \, \frac{\omega}{c}   
\left(\chi' \Gamma' +\chi'' \Gamma''\right)
|{\cal A}(z)|^2,
\EEq
and differential absorbance (which generalizes the concept of circular 
dichroism (CD), {\it cf}.\ Eq.\ (7) in \cite{yin2013interpreting}),
\BEq
{\rm CD} 
\sim
\Delta A(z)
\equiv
-\frac{\omega d}{c} \, \frac{2 \omega}{c} 
\left( 
\Gamma'' \, {\cal A}'(z) + \Gamma' \, {\cal A}''(z)
\right),
\EEq
where we used primes to indicate the real and imaginary parts of various 
quantities involved,
$\chi = \chi'+i\chi''$,
$\Gamma = \Gamma'+i\Gamma''$,
${\cal A} = {\cal A}'+i{\cal A}''$.
It is then straightforward to show using simple algebra that (unnormalized) 
transmitted differential intensity is given by 
\BEq
\label{eq:deltaTlaser}
\Delta T(z)
\equiv
|{\cal T}_{{\rm L}}(z)|^2-|{\cal T}_{{\rm R}}(z)|^2
=
\Delta B(z) + \Delta A(z),
\EEq
which is plotted for a laser beam of finite width in Fig.\ \ref{fig:14}.

\begin{figure} [!ht]
\includegraphics[angle=0,width=0.6\linewidth]{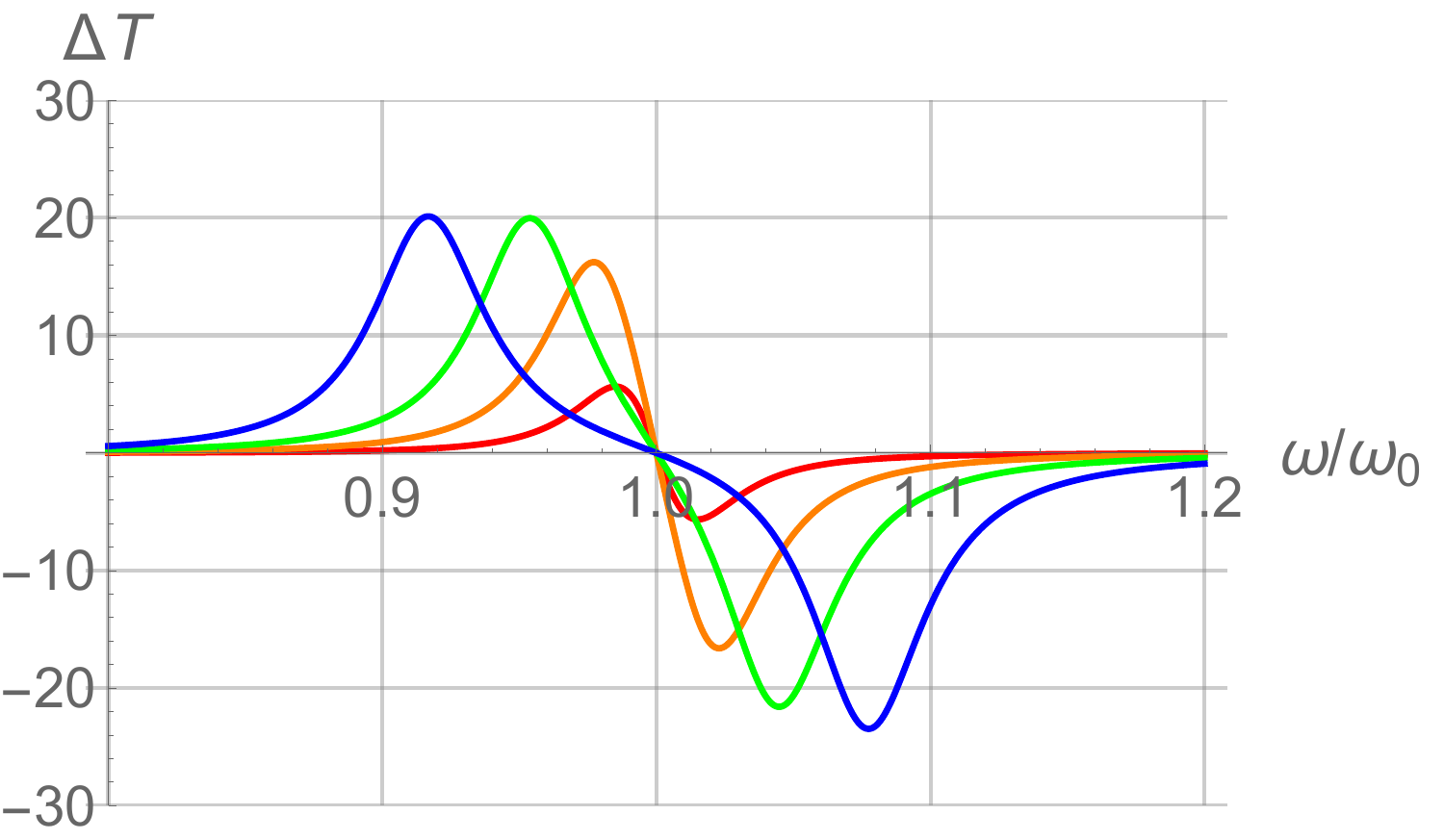}
\caption{ \label{fig:15}  
Transmitted differential intensity, $\Delta T$ (unnormalized), 
as predicted by Eq.\ (\ref{eq:deltaTwide}),
for a wide beam ($b=\infty$, ${\cal A}(z) = 1$) passing through a 
$C_4$-symmetric BK plate.
Here, $kd=0.1$, $\omega_{\rm p}/\omega_0=2$,
$\gamma/\omega_0=0.05$, $\omega_{\rm c}/\omega_0=$ 0.1 (red), 
0.2 (orange), 0.3 (green), 0.4 (blue).
}
\end{figure}

For a wide beam (${\cal A} = 1$) this becomes $z$-independent,
\BEq
\label{eq:deltaTwide}
\Delta T
=
\frac{\omega^2 d^2}{c^2} \, 
\frac{16 \, \omega_{\rm p}^2 \, \omega_{\rm c}^2   
\left(\omega_{0}^2-\omega^2\right) 
\left(\omega_{\rm p}^2 \left({\omega} d/c \right) - {\gamma}\omega \right)
}{
\left(\left(\omega^2-\omega_{0}^2\right)^2-\omega_{\rm c}^4\right)^2
+
2 \, {\gamma}^2 \omega^2 
\left(\left(\omega^2-\omega_{0}^2\right)^2+\omega_{\rm c}^4\right)
+{\gamma}^4 \omega^4
}\,,
\EEq
as shown in Fig.\ \ref{fig:15}

Finally, for an incoming beam {\it linearly} polarized along the $x$-axis, with 
${\cal E}_{x}={\cal E}_{0}$, ${\cal E}_{y}=0$, the corresponding 
transmitted field is given by 
\begin{align}
E_x(z,t)
=
\left(
1
+ 
i\,\frac{\omega \chi d}{2c}\, {\cal A}(z)
\right)
{\cal E}_{0} \,  e^{-i\omega (t-z/c)},
\quad
E_y(z,t)
&=
-
\frac{\omega^2 \Gamma d}{2c^2} \, 
 {\cal A}(z) \, {\cal E}_{0}\,  e^{-i\omega (t-z/c)}.
\end{align}
If we work with a wide beam ($b=\infty$, ${\cal A} = 1$), assume no 
losses ($\gamma = 0$), and take $\omega$ to be far from each of the 
two resonances, $\omega_{\pm}=\sqrt{\omega_0^2\pm \omega_{\rm c}^2}$, 
then the $\chi$-dependent term can be dropped and we get the standard 
expression for polarization rotation angle, 
\BEq
\psi \approx - \frac{\omega^2}{2c^2} \, \Gamma d \,,
\EEq
where the minus sign is due to our left-handed choice for the chirality of 
the BK plate.

\section{Summary}

We developed a simple physically motivated theoretical method 
for the analysis of electromagnetic processes in nanorod-based 
plasmonic metamaterials that is not rooted in macroscopic electrodynamics. 
Specific predictions were made for a single sheet of aligned nanorods and 
for a chiral plate made of corner-stacked nanorods, which can easily be 
extended to more complicated multi-stacked structures consisting of 
dipole-like basic elements. We proposed several interference experiments 
to test our theory that can be performed using readily available equipment. 
Due to high sensitivity of the predicted interference patterns to small changes 
in various system's parameters (such as, e.\ g., surface density of the deposited 
nanorods or resonant frequencies of individual nanorods) the proposed 
experimental implementations could potentially be used in the development 
of quality standards for various metamaterial devices. 
In principle, our approach should also be able to handle systems that exhibit 
nonlinear and higher-order multipole behavior, though the mathematical 
details accompanying such generalization may likely become rather complicated.

\appendix

\section{Elementary theory of scattering on a plane of charges}
\label{appendix:1}

Here for the sake of completeness we re-derive Eqs.\ 
(\ref{eq:ExIMPLEMENTATIONII}) and (\ref{eq:ExTransmittedChiral}), 
(\ref{eq:EyTransmittedChiral}) in the case of a wide beam 
($b=\infty$, ${\cal A} =1$) using a more traditional approach to scattering.
Our discussion parallels that of Ref.\ \cite{chiu1976infrared}.

For harmonically oscillating fields, $\bm{E}$, $\bm{B}$,
charges, $\rho$, and currents, $\bm{j}$,
Maxwell's equations take the form (Coulomb's constant dropped, 
the magnetic field $\bm{B}$ is re-scaled by the speed of light $c$),
\begin{align}
\label{eq:ME01omega}
\nabla \cdot \bm{E}(\bm{r},\omega)  &= 4\pi \rho(\bm{r},\omega), 
\\
\label{eq:ME02omega}
\nabla \times \bm{E}(\bm{r},\omega) &= \frac{i\omega}{c}\bm{B}(\bm{r},\omega), 
\\
\label{eq:ME03omega}
\nabla \cdot \bm{B}(\bm{r},\omega) &= 0, 
\\
\label{eq:ME04omega}
\nabla \times \bm{B}(\bm{r},\omega) 
&= 
-\frac{i\omega}{c}\bm{E}(\bm{r},\omega)
+ \frac{4\pi}{c} \bm{j}(\bm{r},\omega).
\end{align}
If we assume a scenario in which a plane electromagnetic wave propagating in the positive $z$-direction strikes at normal incidence infinitesimally thin (essentially, $\delta$-function like) slab of some material located at $z=0$, then for points outside of the $xy$-plane the solution is
\begin{align}
\bm{E}(z<0,\omega) 
= 
\begin{pmatrix}
{\cal E}_{x}^{(i)}\\
{\cal E}_{y}^{(i)}\\
0
\end{pmatrix}
e^{-i(\omega t - k z)}
+
\begin{pmatrix}
{\cal E}_{x}^{(r)}\\
{\cal E}_{y}^{(r)}\\
0
\end{pmatrix}
e^{-i(\omega t + k z)},
\quad
\bm{E}(z>0,\omega) 
= 
\begin{pmatrix}
{\cal E}_{x}^{(t)}\\
{\cal E}_{y}^{(t)}\\
0
\end{pmatrix}
e^{-i(\omega t - k z)},
\end{align}
and similarly for the magnetic field,
with
\begin{align}
&{\cal B}_{x}^{(i)}=-{\cal E}_{y}^{(i)},
\quad
{\cal B}_{y}^{(i)}={\cal E}_{x}^{(i)},
\\
&{\cal B}_{x}^{(r)}={\cal E}_{y}^{(r)},
\quad
{\cal B}_{y}^{(r)}=-{\cal E}_{x}^{(r)},
\\
&{\cal B}_{x}^{(t)}=-{\cal E}_{y}^{(t)},
\quad
{\cal B}_{y}^{(t)}={\cal E}_{x}^{(t)},
\end{align}
and the boundary conditions at the interface,
\begin{align}
& {\cal E}_{x}^{(i)}+{\cal E}_{x}^{(r)}={\cal E}_{x}^{(t)},
\quad
{\cal E}_{y}^{(i)}+{\cal E}_{y}^{(r)}={\cal E}_{y}^{(t)},
\\
&{\cal B}_{x}^{(t)}-({\cal B}_{x}^{(i)}+{\cal B}_{x}^{(r)})
=\frac{4\pi}{c}{\cal K}_{y},
\quad
{\cal B}_{y}^{(t)}-({\cal B}_{y}^{(i)}+{\cal B}_{y}^{(r)})
=-\frac{4\pi}{c}{\cal K}_{x},
\end{align}
where ${\cal E}^{(i,r,t)}$ and ${\cal B}^{(i,r,t)}$ are the incident, reflected, and 
transmitted amplitudes, respectively, and $({\cal K}_{x},{\cal K}_{y})$ is the bound 
surface current density. Combining all these relations gives the boundary conditions 
in terms of the electric field only,
\begin{align}
{\cal E}_{x}^{(i)}-{\cal E}_{x}^{(r)}-{\cal E}_{x}^{(t)}
=\frac{4\pi}{c}{\cal K}_{x},
\quad
{\cal E}_{y}^{(i)}-{\cal E}_{y}^{(r)}-{\cal E}_{y}^{(t)}
=\frac{4\pi}{c}{\cal K}_{y},
\end{align}
which for transmitted amplitudes become
\begin{align}
\label{eq:AtransmissionVSincidence}
{\cal E}_{x}^{(t)}={\cal E}_{x}^{(i)}-\frac{2\pi}{c}{\cal K}_{x},
\quad
{\cal E}_{y}^{(t)}={\cal E}_{y}^{(i)}-\frac{2\pi}{c}{\cal K}_{y}.
\end{align}

For a single sheet of aligned nanorods, we get on the basis of Eq.\ (\ref{eq:xyPositions}),
\BEq
\label{eq:xyVelocities}
    \dot{x}(t) = -i\omega x_0  e^{-i\omega t},
\quad
x_0
=
\frac{q}{m \Omega^2}{\cal E}_{x}^{(i)},
\quad
\dot{y}(t)=0,
\EEq
and
\BEq
{\cal K}_{x}
=
\eta q \dot{x}(t) 
=
-i\omega \frac{\eta q^2}{m D} \frac{D}{\Omega^2}{\cal E}_{x}^{(i)},
\quad
{\cal K}_{y}
= 0,
\EEq
which upon restoring Coulomb's constant recovers 
Eq.\ (\ref{eq:ExIMPLEMENTATIONII}) with ${\cal A}=1$ (wide beam). 
Notice that in our derivation we assumed that the nanorods were not densely 
packed and were driven by the {\it local} field, which, following standard 
practice \cite{schaferling2017chiral, svirko2000polarization}, 
we took to be ${\cal E}_{x}^{(i)}$. 
This is in contrast with the tightly packed scenario which would typically be 
handled using macroscopic electrodynamics by taking
\BEq
{\cal K}_{x} = \sigma_x  {\cal E}_{x}^{(t)},
\EEq
where $\sigma_x $ is the conductivity, with the result 
(see Eq.\ (1) in \cite{mikhailov2004microwave}), 
\BEq
{\cal E}_{x}^{(t)}
=
\frac{{\cal E}_{x}^{(i)}}{1+2\pi \sigma_x /c}.
\EEq
The tightly packed scenario clearly allows the possibility of zero 
transmission for very large $\sigma_x$
(think of aluminum foil reflecting all incident light, for example). 
In our theory that situation would roughly correspond to
\BEq
\label{eq:xyVelocitiesTIGHT}
    \dot{x}(t) = -i\omega x_0  e^{-i\omega t},
\quad
x_0
=
\frac{q}{m \Omega^2}{\cal E}_{x}^{(t)},
\quad
\dot{y}(t)=0,
\EEq
[compare with Eq.\ (\ref{eq:xyVelocities}) above], which after restoring Coulomb's constant 
would give,
\begin{align}
{\cal T}(z)
\equiv
\frac{{\cal E}_{x}^{(t)}}{{\cal E}_{x}^{(i)}}
=
\frac{1}
{1 - i \, \frac{\omega D}{2c} \frac{\omega_{{\rm p}1}^2}{\Omega^2}},
\end{align}
and, in first order,
\begin{align} 
{\cal T}(z) 
\approx
1 +i \, \frac{\omega D}{2c} \frac{\omega_{{\rm p}1}^2}{\Omega^2},
\end{align}
in agreement with Eq.\ (\ref{eq:ExIMPLEMENTATIONII}).

For a chiral plate viewed as two planes of orthogonally oscillating charges, the first set of 
boundary conditions is
\begin{align}
\label{eq:AtransmissionVSincidence1}
{\cal E}_{x}^{(t)}={\cal E}_{x}^{(i)}-\frac{2\pi}{c}{\cal K}_{x},
\quad
{\cal E}_{y}^{(t)}={\cal E}_{y}^{(i)},
\end{align}
while the second is shifted in phase,
\begin{align}
\label{eq:AtransmissionVSincidence2}
{\cal E}_{x}^{(t)}e^{ikd}={\cal E}_{x}^{(i)}e^{ikd},
\quad
{\cal E}_{y}^{(t)}e^{ikd}={\cal E}_{y}^{(i)}e^{ikd}-\frac{2\pi}{c}{\cal K}_{y},
\end{align}
which on the basis of (\ref{eq:xyPositionsChiral}) and (\ref{eq:xySolutionsChiral}) give
\BEq
{\cal K}_{x}
=
-i\omega \, \frac{\eta q^2}{md} 
\, \frac{\Omega^2 d}{\Omega^4-\omega_{\rm c}^4}
\left(
{\cal E}_{x}^{(i)}-\frac{\omega_{\rm c}^2}{\Omega^2}e^{ikd}{\cal E}_{y}^{(i)}
\right),
\quad
{\cal K}_{y}
=
-i\omega \, \frac{\eta q^2}{md} 
\, \frac{\Omega^2 d}{\Omega^4-\omega_{\rm c}^4}
\left(
{\cal E}_{y}^{(i)}-\frac{\omega_{\rm c}^2}{\Omega^2}e^{-ikd}{\cal E}_{x}^{(i)}
\right) e^{ikd},
\EEq
recovering Eqs.\ (\ref{eq:ExTransmittedChiral}) and (\ref{eq:EyTransmittedChiral})
in the case of a wide beam.

\end{document}